# Stimulated Emission of Radiation in a Single Mode for both Resonance and Non-resonance for Various Initial Photon Distributions.

M. T. Tavis and F. W. Cummings*

## Abstract

This paper reexamines the results of Cummings[1] in which the quantum mechanical two-level-system (TLS) interacts with the electromagnetic field with various initial distributions and extends that work for both resonant and non-resonant to times large enough to display multiple photon echoes. The results presented here include the initial pure coherent state, the field whose initial density matrix is the Gaussian superposition of coherent states (blackbody radiation) and density matrices of the field represented by various combinations of mixed coherent and thermal states with and without squeezing This paper provides, in addition to the matrix elements to the various states, both the algebraic and graphical representation for the first order correlation function $G^{(1)} = \langle E^{-}E^{+}\rangle$ for resonance and non-resonance. It is found that in all case, the application of non-resonance leads to oscillations in the first order correlation which was thought only to apply for the coherent state even for the case of the pure thermal state.

## Historical Background

In early 1966, Cummings posed two questions. One, determine if there was a photon density matrix which describes the mixture of coherent and incoherent states from which the correlation function could be derived and second was it possible to extend his formulation to more than one two level molecule (TLM). The first author was able to, independent of the Lachs' paper,[2] determine that the density matrix could be expressed in the form of a Confluent Hypergeometric function although the alternate representation in terms of the Laguerre polynomial was not found until later. At the same time, the time evolution of the first order correlation function was carried out to times longer than that used in Ref. 1 and the beginnings of the collapse and revival of quantum oscillations was seen. This was published in a paper by Eberly.[3] This author focused on the second question which led to development of the model Hamiltonian for an N 2-Level Molecules with a Quantized Radiation Field Hamiltonian.[4] The intervening years have led to numerous publications with far too many authors to be referenced here; however, the reader may wish to refer to the work referenced herein which contain numerous citations. There does appear to be continuing interest in this model as illustrated by the recent publication by Karatsuba[5] and thus a reason for further investigation.

---

* Michael Tavis (retired), F. W. Cummings (Professor Emeritus, University of California, Riverside)

[1] Cummings, Physical Review, 140, #4A, p1051, 1965.

[2] Gerald Lachs, Theoretical Aspects of Mixtures of Thermal and Coherent Radiation,(1965), Phys. Rev. Vol. 138, #4B, p1012.

[3] J. H. Eberly, N. B. Narozhny, and J.J. Sanchez-Mondrgon, Periodic Spontaneous Collapse and Revival in a Simple Quantum Model, Phys. Rev. Lett. 44, 1323-1326 (1980)

[4] Tavis, A Study of an N Molecule-Quantized Radiation Field–Hamiltonian, Thesis 1968 also in arXiv.org 1206.0078 June 1, 2012

## Algebraic Preliminaries

Reference (1) provides the required basic formulations. The first order correlation function for the resonant case assuming that the TLS is initially in the upper state is given by

$$\langle E^{-}E^{+}\rangle(t) = \left(\frac{2\gamma}{\mu}\right)^{2}[\bar{n} + S_1(\bar{n},\gamma t)]$$
$$S_1(\bar{n},\gamma t)_r = \sum_{n=0}^{\infty} \rho_{nn} Sin^2(n+1)^{1/2}\gamma t. \qquad 1$$

In expression (1), $\rho_{nn}$ are the diagonal elements of the photon density matrix, $\mu$ is the magnitude of the electric-dipole-moment, $\gamma$ is the magnitude of the interaction constant in the interaction energy associated with the interaction between the TLS and the photon field and $\bar{n}$ is the mean number of photons. Below we will plot the second equation in the above expression and note that the subscript r in that expression stands for the resonant case. Note that the negative value of $S_1(\bar{n},\gamma t)$ is also the TLS inversion. In the case of non-resonance, the expression above is easily generalized to

$$S_1(\bar{n},\gamma t)_{nr} = \sum_{n=0}^{\infty} \rho_{nn} \frac{(n+1)\gamma^2}{{\beta_{n+1}}^2} Sin^2\beta_{n+1}t, \qquad 2$$

where ${\beta_n}^2 = \frac{1}{4}[(\omega - \Omega)^2 + 4n\gamma^2]$. If we define $\Delta = \frac{(\boldsymbol{\omega}-\boldsymbol{\Omega})^2}{\boldsymbol{4\gamma^2}}$, then Eq. 2 can be rewritten by using the photon revival time found by Eberly[3] ($\tau_R = 2\pi\frac{\sqrt{\bar{n}+1+\Delta}}{\gamma}$) as

$$S_1(\bar{n},\gamma t)_{nr} = \sum_{n=0}^{\infty} \rho_{nn} \frac{(n+1)}{n+1+\Delta} Sin^2\left[2\,\pi\big((\bar{n}+1+\Delta)(n+1+\Delta)\big)^{1/2}t\right], \qquad 3$$

where $t$ is the dimensionless value of time that is used to plot the figures, which are presented in the following section. Note that the same renormalized time will be used for $S_1(\bar{n},\gamma t)_r$ below. The diagonal elements for the various density matrices as well as the mean photon number and variance are also presented in the following table.

| Coherent State[6] | $\rho_{nn} = \frac{e^{-\beta^2}\lvert\beta\rvert^{2n}}{n!}, \bar{n} = \lvert\beta\rvert^2, var = \lvert\beta\rvert^2$ | 4 |
|---|---|---|
| Thermal State[(6)] | $\rho_{nn} = \frac{1}{\bar{n}_T + 1}\left(\frac{\bar{n}_T}{\bar{n}_T + 1}\right)^n, \bar{n} = \bar{n}_T, var = {\bar{n}_T}^2 + \bar{n}_T$ | 5 |
| Fock State | $\rho_{nm} = \delta_{nm},\ \ \bar{n} = n, var = 0$ | 6 |

---

[5] Anatolii A Karatsuba and Ekatherina A Karatsuba, A resummation formula for collapse and revival in the Jaynes-Cummings model, J. Phys. A: Math. Theor. 42 (2009), 195304

[6] Roy J. Glauber, Coherent and Incoherent States of the Radiation Field, Physical Review, Vol. 131, #6, Sept. 1963, p. 2766.

|  |  |  |
|---|---|---|
| Mixed Coherent and Thermal State[2] | $\rho_{nm} = \frac{e^{-\frac{\beta^2}{\bar{n}_T+1}}}{(\bar{n}_T+1)}\left(\frac{\bar{n}_T}{\bar{n}_T+1}\right)^n M\left(-n,1,\frac{-\beta^2}{\bar{n}_T(\bar{n}_T+1)}\right),$ <br> $\bar{n} = \vert\beta\vert^2 + \bar{n}_T, \qquad var = \vert\beta\vert^2(1+2\bar{n}_T) + \bar{n}_T{}^2 + \bar{n}_T$ | 7 |
| Squeezed Vacuum State[7] | $\rho_{nm} = \frac{\left[\frac{1}{2}Tanhr\right]^n n!}{Coshr\left(\frac{n}{2}!\right)^2} \quad n\ even, \qquad \bar{n} = Sinh^2 r$ <br> $= 0 \qquad n\ odd, \qquad var = \frac{Sinh^2(2r)}{2}$ | 8 |
| Squeezed Fock State[7] for state $l$ | $\rho_{nn}(l) = \frac{l!\,n!}{(Coshr)^{2n+1}}\left(\frac{1}{2}Tanhr\right)^{l-n} S(r,n,l) \ \vert n-l\vert\ even, \qquad \bar{n} = l + (2l+1)Sinh^2 r$ <br> $= 0 \qquad \vert \mathrm{n}-\mathrm{l}\vert\ \mathrm{odd}, \qquad var = \frac{1}{2}(l^2+l+1)Sinh^2(2r)$ | 9 |
| Squeezed Thermal State[7] | $\rho_{nn} = \frac{1}{\bar{n}_T+1}\sum_{l=0}^{\infty}\rho_{nn}(l)\left(\frac{\bar{n}_T}{\bar{n}_T+1}\right)^l,$ <br> $\bar{n} = \bar{n}_T + (2\bar{n}_T+1)Sinh^2 r\,, var = -\frac{1}{4} + \left(\bar{n}_T + \frac{1}{2}\right)^2 Cosh(4r)$ | 10 |
| Squeezed Coherent State[8] | $\rho_{nn} = N_1(\beta,r,\psi)\frac{\left[\frac{1}{2}Tanh(r)\right]^n}{n!}\left\vert H_n\left\{\frac{\vert\beta\vert}{\sqrt{2}}\left[e^{-i\frac{\psi+\pi}{2}}Tanh^{-\frac{1}{2}}(r) + e^{i\frac{\psi+\pi}{2}}Tanh^{\frac{1}{2}}(r)\right]\right\}\right\vert^2$ <br> $\bar{n} = Sinh^2 r + \vert\beta\vert^2, var = \vert\beta\vert^2[Cosh(2r) + Cos(\psi)Sinh(2r)] + \frac{Sinh^2(2r)}{2}$ | 11 |
| Mixed Squeezed Coherent State and Thermal State[9] | $\rho_{nn} = N_2(\beta,\bar{n}_T,r)\frac{1}{n!}\left[\frac{\bar{n}_T}{1+\bar{n}_T}\right]^n H_{n,n}(r_1,r_2)$ <br> $\bar{n} = Sinh^2 r + \vert\beta\vert^2 + \bar{n}_T$ <br> $var = \vert\beta\vert^2[Cosh(2r) + Cos(\psi)Sinh(2r)] + \frac{Sinh^2(2r)}{2} + 2\bar{n}_T(Sinh^2 r + \vert\beta\vert^2) + \bar{n}_T{}^2 + \bar{n}_T$ | 12 |
| Displaced Squeezed Thermal State (DSTS)[10] | $\rho_{nn} = \pi Q(0)\tilde{A}^n\sum_{q=0}^{n}\frac{1}{q!}\binom{n}{q}\left(\frac{\lfloor\tilde{B}\rfloor}{2\tilde{A}}\right)^q\left\vert H_q\left(\frac{\tilde{C}}{\sqrt{2\tilde{B}}}\right)\right\vert^2$ <br> $\bar{n} = \bar{n}_T + (2\bar{n}_T+1)Sinh^2 r + \vert\beta\vert^2,$ <br> $var = -\frac{1}{4} + \vert\beta\vert^2(1+2\bar{n}_T)[Cosh(2r) + Cos(\psi)Sinh(2r)] + \left(\bar{n}_T + \frac{1}{2}\right)^2 Cosh(4r)$ | 13 |
| Displaced Number (Fock) State[11] for state $l$ | $\rho_n(l) = \begin{cases} \frac{n!}{l!}\vert\beta\vert^{2(l-n)}e^{-\vert\beta\vert^2}\vert\mathcal{L}_n^{l-n}\vert^2 = \frac{\vert\beta\vert^{2(l-n)}e^{-\vert\beta\vert^2}}{n!\,l!}\left\vert\sum_{k=0}^{n}\frac{n!\,l!\,(-1)^k\vert\beta\vert^{2(n-k)}}{k!\,(n-k)!\,(l-k)!}\right\vert^2 & l \geq n \\ \frac{l!}{n!}\vert\beta\vert^{2(n-l)}e^{-\vert\beta\vert^2}\left\vert\mathcal{L}_l^{n-l}\right\vert^2 = \frac{\vert\beta\vert^{2(n-l)}e^{-\vert\beta\vert^2}}{n!\,l!}\left\vert\sum_{k=0}^{l}\frac{n!\,l!\,(-1)^k\vert\beta\vert^{2(l-k)}}{k!\,(n-k)!\,(l-k)!}\right\vert^2 & n > l \end{cases}$ <br> $\bar{n} = l + \vert\beta\vert^2, \qquad var = \overline{n^2} - \bar{n}^2 = (2l+1)\vert\beta\vert^2$ | 14 |

[7] M. S. Kim, F. A. M. de Oliveira, and P. L. Knight; Properties of squeezed number states and squeezed thermal states; Physical Review A, Vol. 40, # 5, Sept. 1, 1989

[8] J. J. Gong and P. K. Aravind, Expansion coefficients of a squeezed coherent state in the number state basis, The American Journal of Physics, Vol. 58, Issue 10 Oct. 1990

[9] A. Vourdas, Superposition of Squeezed Coherent States with Thermal Light, Phy. Rev A Vol. 34, #4, Oct. 1986, p. 2366. Note that the expressions for the mean and variance were not given in this reference. In fact, the authors have not found this expression for this case in any reference.

[10] Paulina Marian and Tudor A. Marian, Squeezed States with Thermal Noise. I Photon-Number Statistics, Phy. Rev. A, Vol. 47, #5, May 1993, p. 4474.

[11] F. A. M. de Oliveira, M. S. Kim, P. L. Knight and V. Bužek, Properties of displaced number states, Phy. Rev. A, Vol. 41, #5, p2645

| Squeezed Displaced Number State[12] for initial state m | $\rho_{nn} = \lvert\langle n\rvert\beta,m\rangle_g\rvert^2 = \frac{n!}{m!\,Cosh(r)}\left[\frac{Tanh(r)}{2}\right]^{m+n} \times Exp\{-\lvert\beta\rvert^2[1-Cos(\psi)Tanh(r)]\}\left\lvert\sum_{i=0}^{min(m,n)}\frac{\binom{m}{i}}{(n-i)!}\left[-\frac{4}{Sinh^2(r)}\right]^{\frac{i}{2}} SS(i,m,n,\lvert\beta\rvert^2,r,\psi)\right\rvert^2$ <br> $\bar{n} = \langle n\rangle = \lvert\beta\rvert^2 + (2m+1)Sinh^2(r) + m$ <br> $var = \langle(\Delta n)^2\rangle = \lvert\beta\rvert^2[Cosh(2r)+Cos(\psi)Sinh(2r)](2m+1) + \frac{1}{2}(m^2+m+1)Sinh^2(2r)$ | 15 |
|---|---|---|

In the table above, a consistent set of variables have been used to simplify comparison. $|\beta|^2$ is the mean number of coherent photons, $\bar{n}_T$ is the mean number thermal photons, r is the squeezing parameter and ψ is the phase factor that accounts for the phase of squeezing parameter and the phase of $\beta$. Upon examination of the various expressions for the variance, it is easily seen that the smallest variance occurs for ψ=$\pi$. This is the value chosen for most calculations presented below. In addition, the Confluent Hypergeometric Function can be expressed in terms of the Laguerre Polynomial,

$$M\left(-n,1,\frac{-|\beta|^2}{\bar{n}_T(\bar{n}_T+1)}\right) = L_n^0\left(\frac{-|\beta|^2}{\bar{n}(\bar{n}+1)}\right) \qquad 16$$

In Eq. (9), the quantity $S(r,n,l)$ is given by

$$S(r,n,l) = \left|\sum_m \frac{(-1)^m\left(\frac{Sinhr}{2}\right)^{2m}}{m!\,(n-2m)!\,[m+(l-m)/2]!}\right|^2. \qquad 17$$

Note that in the sum for S, the value of m is limited so that $\frac{1}{2}(n-l) \leq m \leq \frac{n}{2}$ and m must be a non-negative integer. This sum can be expressed in terms of the Hypergeometric function[13]

$$S(r,n,l) = \begin{cases} \left(\frac{Sinhr}{2}\right)^{2n}\left|\frac{F\left(-\frac{l}{2},-\frac{n}{2},\frac{1}{2},-\frac{1}{(Sinhr)^2}\right)}{\left(\frac{n}{2}\right)!\left(\frac{l}{2}\right)!}\right|^2 & l \text{ and } n \text{ even} \\ \left(\frac{Sinhr}{2}\right)^{2(n-1)}\left|\frac{F\left(-\frac{l-1}{2},-\frac{n-1}{2},\frac{3}{2},-\frac{1}{(Sinhr)^2}\right)}{\left(\frac{n-1}{2}\right)!\left(\frac{l-1}{2}\right)!}\right|^2 & l \text{ and } n \text{ odd} \end{cases} \qquad 18$$

The normalization factors in Eq. (11) and (12) are given by

$$N_1(\beta,r,\psi) = \frac{exp\{-|\beta|^2[1-Cos(\psi)Tanh(r)]\}}{\mathrm{Cosh}(r)}, \qquad 19$$

[12] P. Král, Displaced and Squeezed Fock states, Journal of Modern Optics, Vol. 37, #5, p889, 1990.
[13] This was found independently from Paulina Marian, Higher-order squeezing properties and correlation functions for squeezed number states, Phy. Rev A, Vol 44, #5, p.3325.

$$N_2(\beta,\bar{n}_T,r)=\frac{Exp\left[-\frac{\beta^2}{\left[\bar{n}_T+e^{-2r}Cosh^2(r)\left(1+Tanh(r)\right)\right]}\right]}{Cosh(r)\sqrt{\{\bar{n}_T[1+Tanh(r)]+e^{2r}\}\{\bar{n}_T[1-Tanh(r)]+e^{-2r}\}}}$$

In Eq. 12, $H_{n,n}(r_1,r_2)$ are the diagonal elements of the generalized Hermite Polynomial whose expression is given by[14(9)]

$$\begin{gathered}
H_{n,m}(r_1,r_2)=(-1)^{m+n}Exp\left(\frac{1}{2}\sum_{i,j=1}^{2}c_{i,j}r_ir_j\right)\frac{\partial^{m+n}}{\partial r_1^m\partial r_2^n}Exp\left(-\frac{1}{2}\sum_{i,j=1}^{2}c_{i,j}r_ir_j\right)\\
r_1=r_2=-\left[1-\frac{tanh(r)}{K\bar{n}_T}\right]^{-1}e^{-r}\left(\frac{|\beta|^2}{\bar{n}_T}\right)^{1/2}cosh(r)(1+\bar{n}_T)^{-1/2}\\
c_{11}=c_{22}=\frac{1}{2}(1+\bar{n}_T)(\Lambda-K)+tanh(r)\left(1+\frac{1}{\bar{n}_T}\right)\\
c_{12}=c_{21}=-\frac{1}{2}(1+\bar{n}_T)(\Lambda+K)\\
K=\frac{1}{cosh^2(r)\left[\left(1-tanh(r)\right)\bar{n}_T+e^{-2r}\right]}\\
\Lambda=\frac{1}{cosh^2(r)\left[\left(1+tanh(r)\right)\bar{n}_T+e^{2r}\right]}
\end{gathered}\qquad 20$$

In order to perform the calculations needed to produce the figures below, the recursion relations for the generalized Hermite Polynomial are used. Namely

$$\begin{gathered}
H_{0,0}=1,H_{1,0}=c_{11}r_1+c_{12}r_2,H_{0,1}=c_{22}r_2+c_{12}r_1\\
H_{1,1}=(c_{11}r_1+c_{12}r_2)(c_{22}r_2+c_{12}r_1)-c_{12}\\
H_{n+1,m}=(c_{11}r_1+c_{12}r_2)H_{n,m}-nc_{11}H_{n-1,m}-mc_{12}H_{n,m-1}\\
H_{n,m+1}=(c_{22}r_2+c_{12}r_1)H_{n,m}-mc_{22}H_{n,m-1}-nc_{12}H_{n-1,m}.
\end{gathered}\qquad 21$$

The expressions related to Eq. (13), are given by

$$\begin{gathered}
\pi Q(0)=\frac{1}{\sqrt{(1+A)^2-|B|^2}}Exp\left\{-\frac{(1+A)|C|^2+\frac{1}{2}[B(C^*)^2+B^*C^2]}{(1+A)^2-|B|^2}\right\}\\
\tilde{B}^{\frac{1}{2}}=ie^{i(\varphi/2)}\left|\tilde{B}\right|^{1/2}\ and\ \left(\tilde{B}^*\right)^{1/2}=\left(\tilde{B}^{\frac{1}{2}}\right)^*\\
\tilde{A}=\frac{A(1+A)-|B|^2}{(1+A)^2-|B|^2}=\frac{\bar{n}_T(\bar{n}_T+1)}{{\bar{n}_T}^2+\left(\bar{n}_T+\frac{1}{2}\right)[1+cosh(2r)]}\\
\tilde{B}=\frac{B}{(1+A)^2-|B|^2}=-\frac{e^{i\varphi}\left(\bar{n}_T+\frac{1}{2}\right)sinh(2r)}{{\bar{n}_T}^2+\left(\bar{n}_T+\frac{1}{2}\right)[1+cosh(2r)]}\\
\tilde{C}=\frac{(1+A)C+BC^*}{(1+A)^2-|B|^2}=\frac{C\left[\frac{1}{2}+\left(\bar{n}_T+\frac{1}{2}\right)cosh(2r)\right]-C^*e^{i\varphi}\left(\bar{n}_T+\frac{1}{2}\right)sinh(2r)}{{\bar{n}_T}^2+\left(\bar{n}_T+\frac{1}{2}\right)[1+cosh(2r)]}
\end{gathered}\qquad 22$$

[14] A. Erdelyi, Higer Transcendental Functions, McGraw-Hill, 1953, Vol.2, p 283.

$$A = \bar{n}_T + (2\bar{n}_T + 1)(sinhr)^2; B = -(2\bar{n}_T + 1)e^{i\varphi} sinhr\, coshr$$

$$C = \begin{pmatrix} \beta & DSTS \\ \beta coshr + \beta^* e^{i\varphi} sinhr & SDTS \end{pmatrix}$$

The phase factor $\varphi$ combined with the phase for $\beta$ yields the phase factor $\psi$ discussed above. Again, the recursion relationship for the Hermite Polynomial is used for calculations used to generate the figures below, namely

$$H_{n+1}(z) = 2\, z\, H_n(z) - 2\, nH_{n-1}(z){:}\; H_0(z) = 1, H_1(z) = 2\, z. \qquad 23$$

It should be noted that Eqs. 4, 5, 7, 8, 10 and 11 are all special cases of Eq. 13. This author has not been able to express Eq. 12 as a special case of Eq. 13. Marian[10] has speculated that there is such a relationship but did not express it. In fact, in that reference, they claim that if Vourdas[9] had carried out their calculations to larger values of the squeezing parameter, then they would have seen oscillations in the density matrix. Our calculations do not confirm this.

Finally, we provide the expression for $SS(i, m, n, |\beta|^2, r, \psi)$ in Eq. 15, namely

$$SS(i, m, n, |\beta|^2, r, \psi) = H_{m-i}\left[\frac{|\beta|}{\sqrt{Sinh(2r)}} e^{-i\frac{\psi-2\pi}{2}}\right] H_{n-i}\left[-\frac{|\beta|\left[Cosh(r)e^{i\left(\frac{\psi-\pi}{2}\right)} + Sinh(r)e^{-i\left(\frac{\psi-\pi}{2}\right)}\right]}{\sqrt{Sinh(2r)}}\right] \qquad 24$$

This is self-consistent with the variable seen within the Hermite Polynomial in Eq. 11 for the Squeezed Displaced Vacuum.

## Numerical Examples

Below, numerical results for the density matrices and the first order correlation function $G^{(1)} = \langle E^- E^+ \rangle$, namely Eqs. 1 and 2 are presented. The results are obtained through the use of Wolfram's Mathematica versions 8 and 12. The parameters in the calculation are $|\beta|^2, \bar{n}_T, r, and\ \Delta$. As stated earlier the phase is set to ψ = $\pi$ to minimize the variance for most cases. This leads to a very large set of possible examples of which only a very limited set is presented. (Note: The original versions of this paper did not use the Eberly revival times, thus the times between revivals were quite Large)

### Coherent State

There are only two parameters of interest, namely the mean number of coherent photons $|\beta|^2$ and the off-resonance parameter $\Delta$. Examples have been presented previously[1,3,5] and also in numerous web publications.

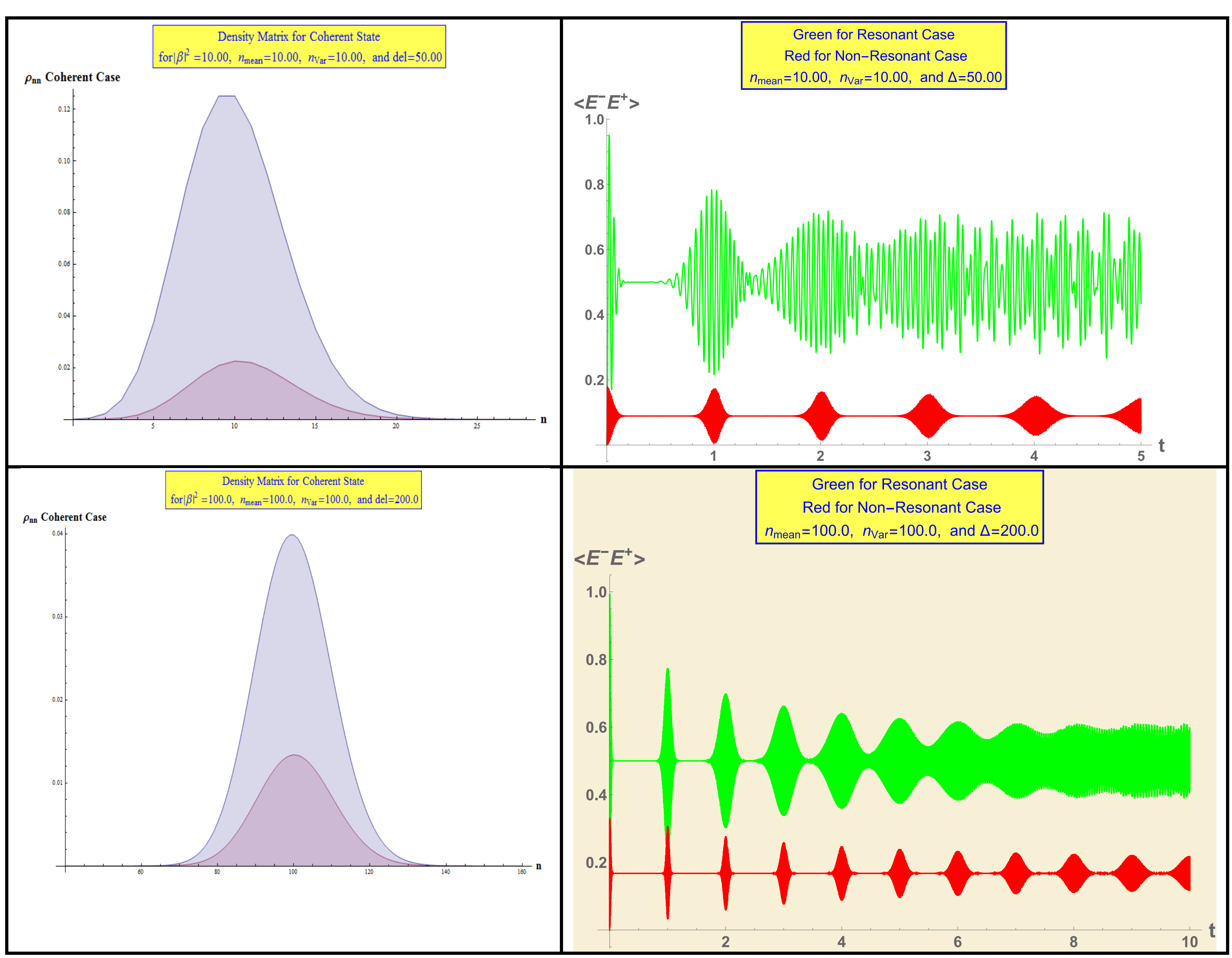

**Table 1: The left-hand columns contain the diagonal elements of the density matrix and for the lower curve the density matrix multiplied by the non-resonance factor in Eq. 2. The right-hand columns contain $S_1(\bar{n},\gamma t)_r$ as the upper curves and $S_1(\bar{n},\gamma t)_{nr}$ for the lower curves**

In the above examples, it is seen that for the resonant case, the decay and revival of oscillation appears to be a function on the mean number of coherent photons with reversion to incoherent oscillation occurring fairly quickly for smaller $|\beta|^2$. In the non-resonant case, the rise and fall of oscillations continues for a much larger time.

## Thermal State

As in the last case, there are two parameters of interest, namely the mean number of thermal photons $\bar{n}_T$ and the off-resonance parameter $\Delta$. The resonant cases were considered by Cummings[(1)].

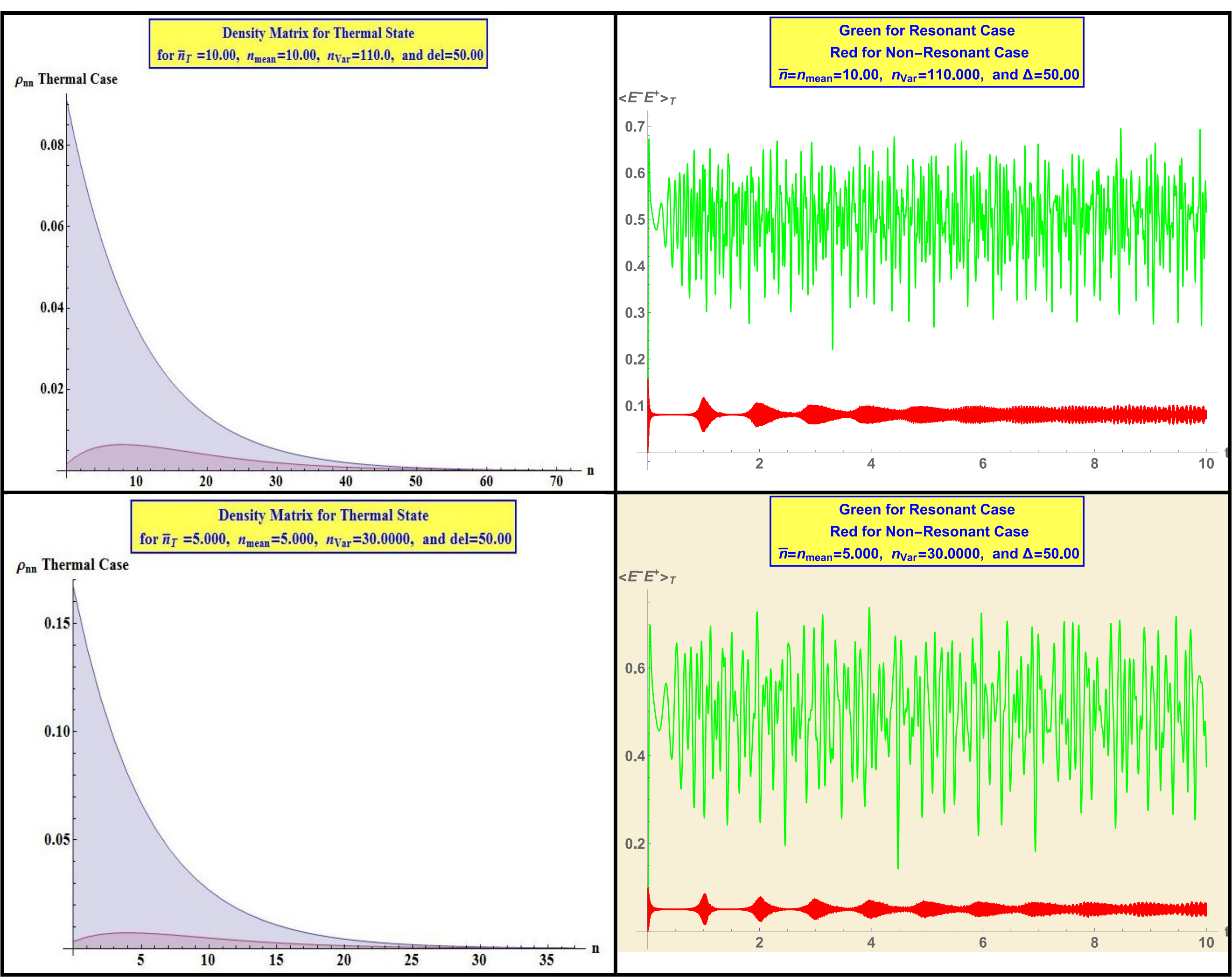

**Table 2. The left-hand columns contain the diagonal elements of the density matrix and for the lower curve the density matrix multiplied by the non-resonance factor in Eq. 2. The right-hand columns contain $S_1(\bar{n},\gamma t)_r$ as the upper curves and $S_1(\bar{n},\gamma t)_{nr}$ for the lower curves**

In is interesting to note that like the coherent case, when non-resonance is applied, the decay and revival of oscillations is seen. The time scale over which the decay and fall are clearly seen is a function of the amount of non-resonance compared to the mean number of photons. In the first example above, the oscillations merge significantly at about the 5th node, while for the lower case nearer the end of the time displayed. In fact, we have calculated the results for $\bar{n}_T$=100 and $\Delta$=200 and found the merge to be almost complete at the third oscillation. For the resonant case, the oscillations are almost completely random and dissimilar to the coherent state.

Second for both the coherent case and the thermal case, the larger the value of non-resonance, $\Delta$, the smaller the inversion. For instance, for $\Delta$ of 10 and $\bar{n}$ of 10, the inversion for the coherent non-resonant case is 25% rather than 50% as in the case of the resonant case. When $\Delta$ is 50 and $\bar{n}$ is 10, the inversion is about 10%. Third, the larger the average number of photons, the larger the inversion up to a maximum of 50%. This is seen for both coherent and incoherent cases. For instance, for the coherent case of $\bar{n}$ of 100, the inversion is 45% at $\Delta$ of 10, 34% at $\Delta$ of 50, and 17% at $\Delta$ of 200. Similarly, for the incoherent case, the inversion is 21% for $\Delta$ of 10 and $\bar{n}$ of 10 while it is 14% for $\Delta$ of 200 and $\bar{n}$ of 100. Finally, and most interestingly, the larger the non-resonance, the more striking and longer lasting the Rabi oscillations. In particular see the case of $\Delta$ = 200 and $\bar{n}$ is 100 for the coherent case. Even for the incoherent case, the effect of the non-resonance is to smooth out the oscillations.

Numerical results will not be presented for the Fock state since the results are uninteresting.

## Mixed Coherent and Thermal State

Numerical results for the mixed coherent and thermal state are presented. The parameters of interest are the mean number of coherent photons$|\beta|^2$, the mean number of thermal photons $\bar{n}_T$ and the off-resonance parameter $\Delta$.

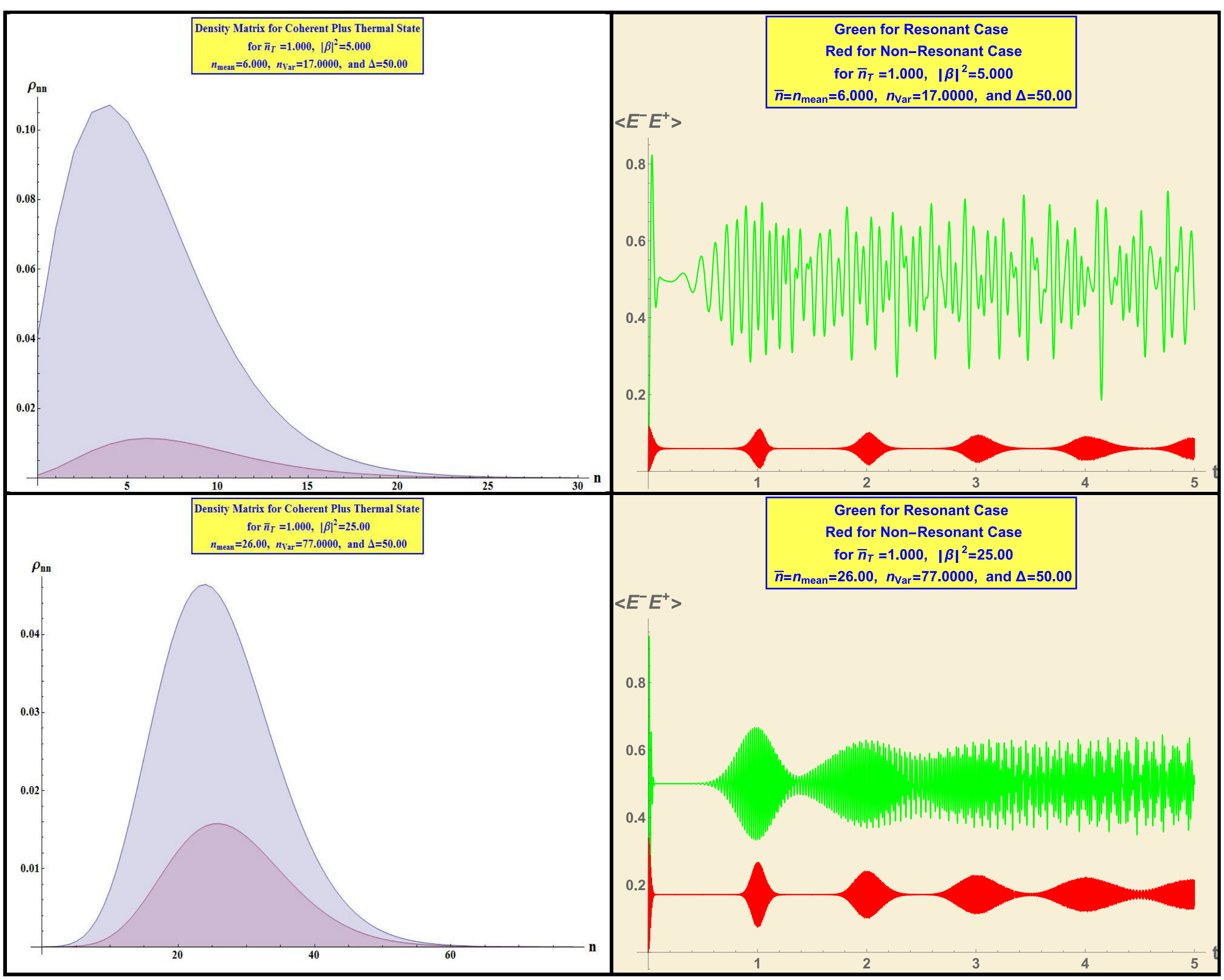

**Table 3.The left-hand columns contain the diagonal elements of the density matrix for the mixed coherent and thermal states and for the lower curve the density matrix multiplied by the no- resonance factor in Eq. 2. The right-hand columns contain the corresponding $S_1(\bar{n}, \gamma t)_r$ as the upper curves and $S_1(\bar{n}, \gamma t)_{nr}$ for the lower curves.**

Some observations can be made based on the calculations for this combination and also hold for the two cases presented above. First note the Rabi like oscillations which occur for the resonant case become more pronounced, the larger the mean number of coherent photons. Second is that these oscillations for the non-resonant case occurs more slowly as Δ grows. For most of the examples shown above, Δ has been set to 50 except for the one coherent case above. If the reader wanted to observe more oscillations, the calculation would have been extended to much larger times which is possible though requiring longer calculation times. Finally, for the immediate case note that the ratio of the mean number of coherent photons to the mean number of thermal photons has a pronounced effect on the resonant case where a ratio smaller than 5 results in the first order correlation function resembling the thermal state much more than the coherent state even for larger values of $|\beta|^2$.

## Squeezed Vacuum State

Numerical results for the squeezed vacuum state are presented. The parameters of interest are the squeezing parameter r and the off-resonance parameter$\Delta$.

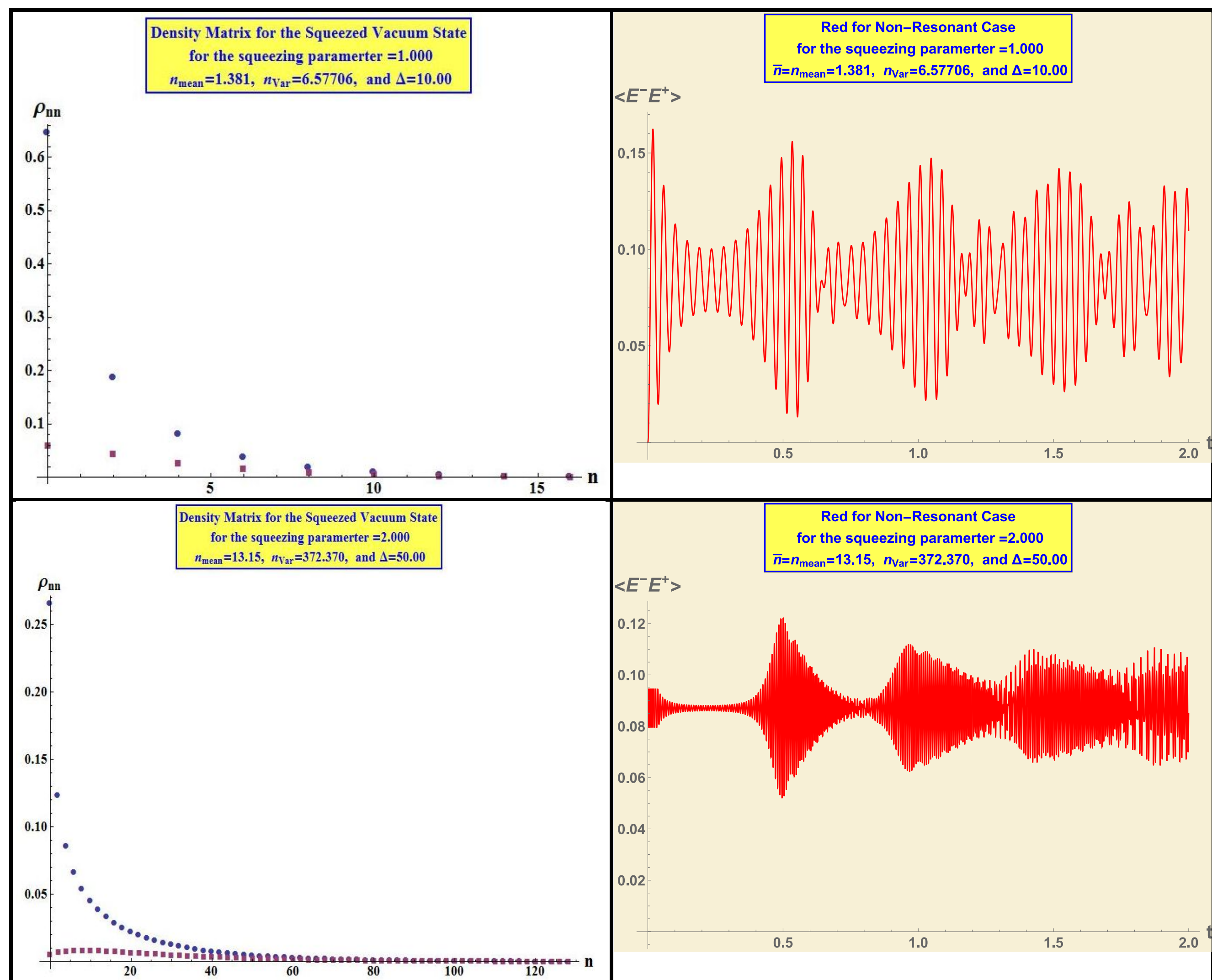

**Table 4: The left-hand columns contain the diagonal elements of the density matrix for the Squeezed Vacuum State and for the lower curve the density matrix multiplied by the non-resonance factor in Eq. 2. The right-hand columns contain the corresponding $S_1(\bar{n},\gamma t)_r$ as the upper curves and $S_1(\bar{n},\gamma t)_{nr}$ for the lower curves. Note that in the second row, $S_1(\bar{n},\gamma t)_r$ has not been plotted since it is of little interest.**

The density matrix for this case appears very similar to the density matrix for the thermal state; however, unlike the thermal case, every other value of the density matrix is zero and the variance is twice the variance of the thermal state for the same mean number of photons. One word of caution in calculating the values of $S_1(\bar{n},\gamma t)_r$ and $S_1(\bar{n},\gamma t)_{nr}$ using Mathematica is that the algorithm used in plotting adjusts the step size over which calculations are performed. This can lead to strange truncations unless care is taken to achieve a small enough step size to exhibit the true behavior of the function. A particularly interesting example exits for a squeezing parameter of 1 and Δ=10. The calculation is only carried out to a normalized time=100.

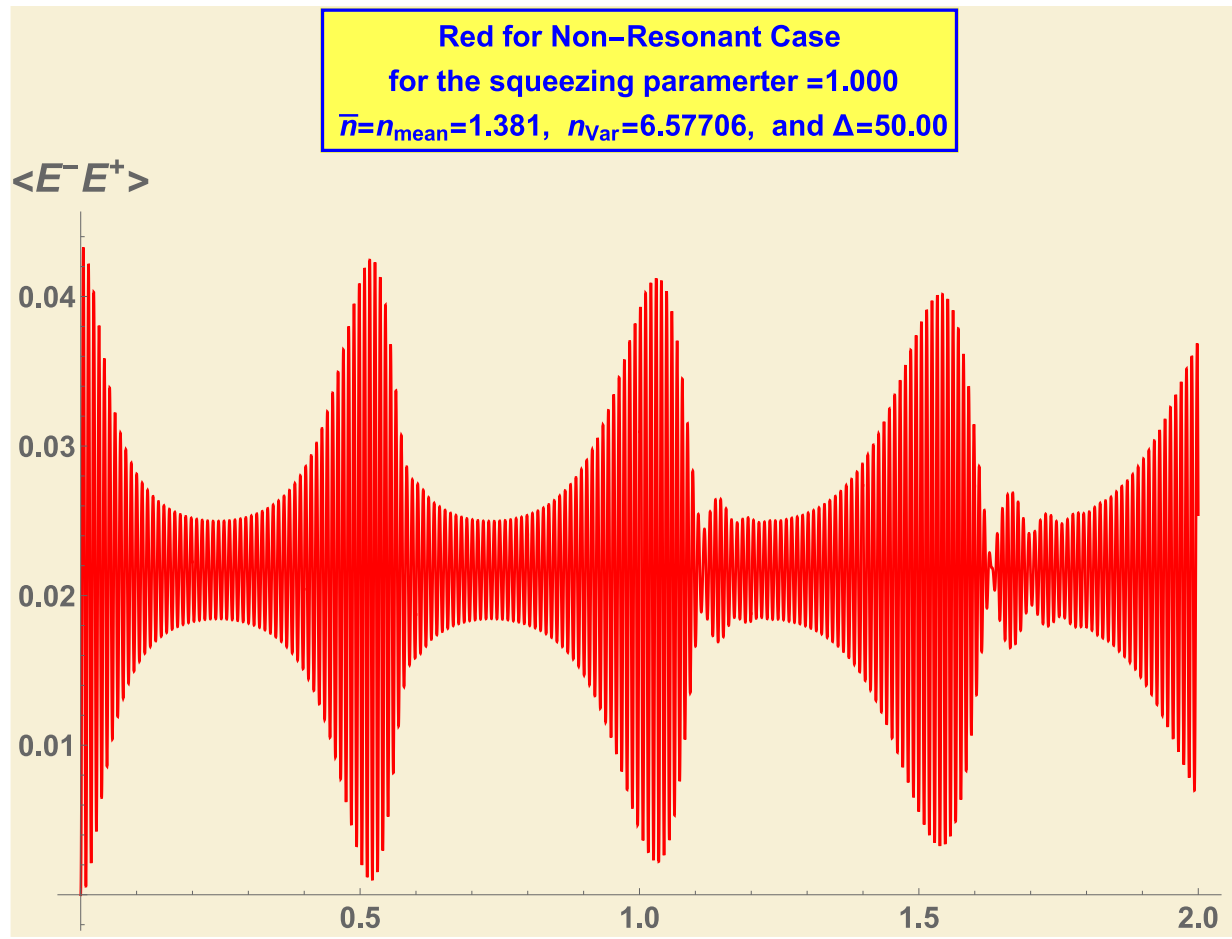

## Squeezed Number State

The density matrix for the squeezed number state is particularly interesting. First, for a moderate number of initial photon number, the variance can be quite large as seen from Eq. 9 for values of squeezing much larger than 1. In addition, the density matrix displays a number of oscillations equal approximately to one half the initial photon number. Several examples are presented below for the density matrix for both even and odd initial photon number and squeezing parameter.

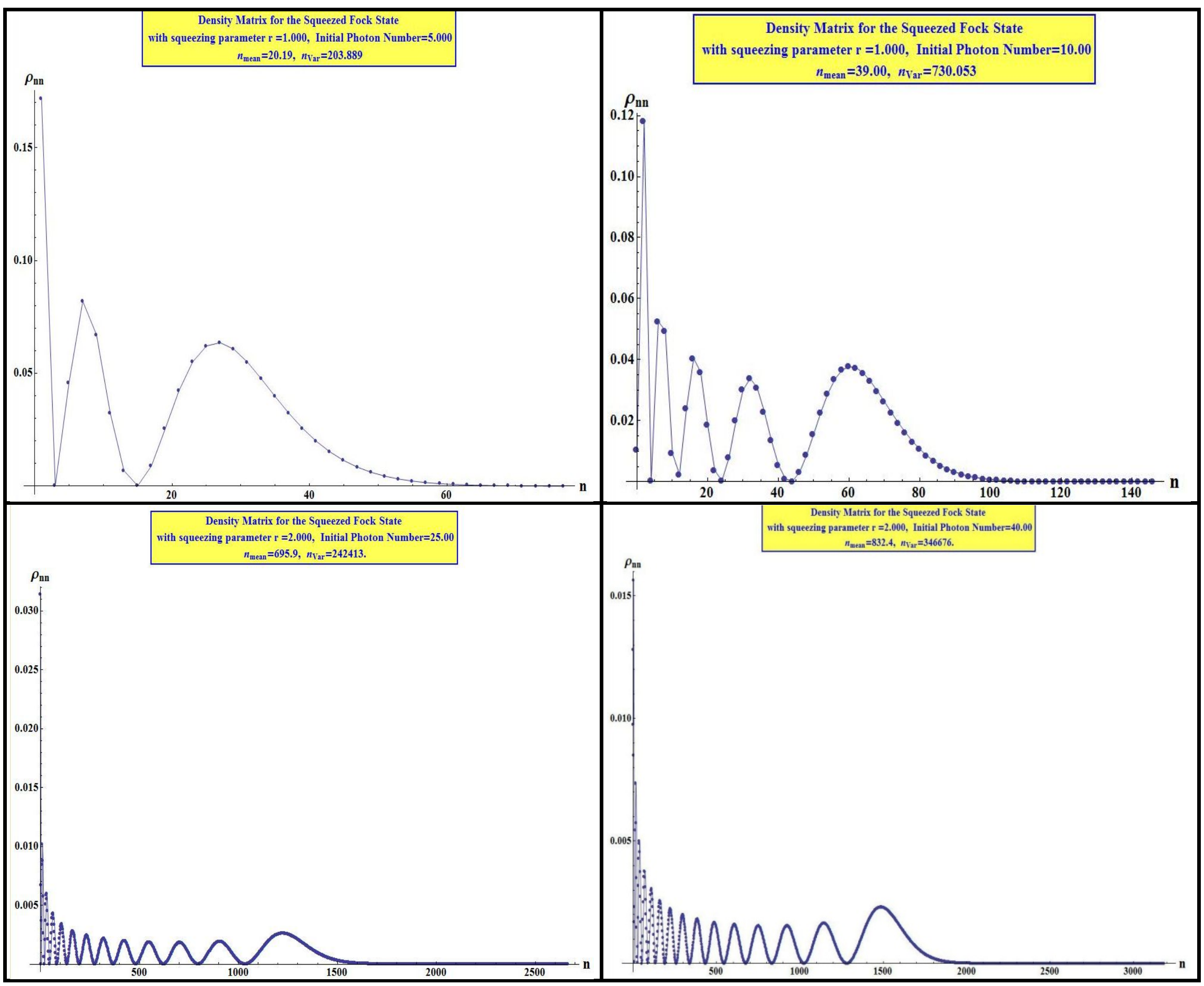

**Table 5: Examples of the density matrix for both even and odd initial photon number and squeezing parameters r=1 and2**

The behavior of the density matrix is consistent with the values of the mean and variance as displayed in Eq. 9 as well as noting  that every other value of the density matrix is zero and depends on the initial photon number being even or odd.

Examples for the time behavior of the first order correlation function for the squeezed number state are presented in the following table.

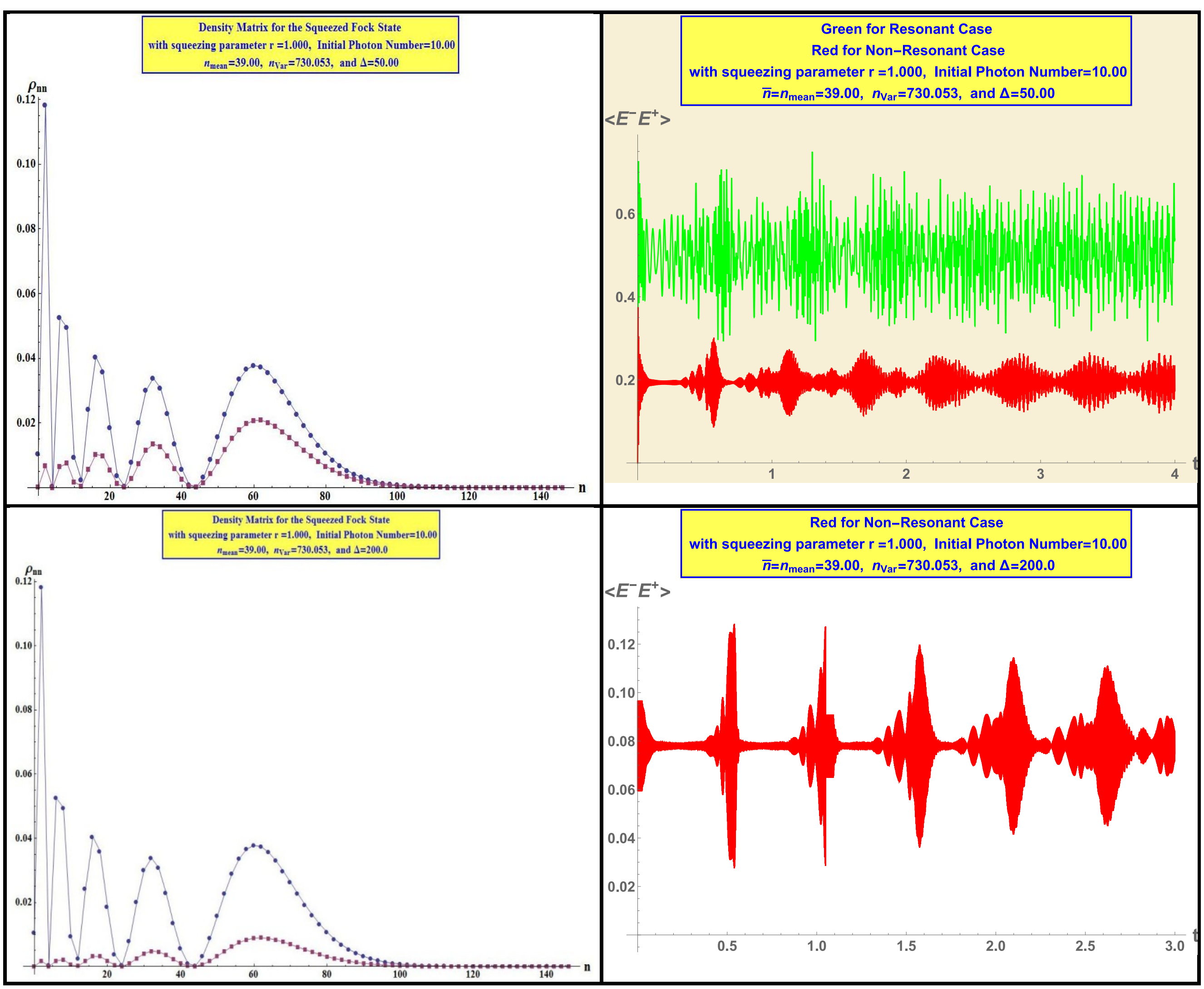

**Table 6: The left-hand columns contain the diagonal elements of the density matrix for the Squeezed Number State and for the lower curve the density matrix multiplied by the non-resonance factor in Eq. 2. The right-hand columns contain the corresponding $S_1(\overline{n}, \gamma t)_r$ as the upper curves and $S_1(\overline{n}, \gamma t)_{nr}$ for the lower curves. Note that $S_1(\overline{n}, \gamma t)_r$ in the second row has not been plotted since it is the same as in the row above.**

As seen in the table above, $S_1(\overline{n}, \gamma t)_{nr}$ displays somewhat the structure of the density matrix. As $\Delta$ increases, some of this structure disappears. This is due to the non resonance factor decreasing the early oscillations in the density matrix in favor for the later ones.

## Squeezed Thermal State

Before proceeding to some numerical examples for this case, it is of interest that Marian[15] has derived an alternate expression for the density matrix in the number representation, namely,

$$P_{Sth}(n) = \frac{(2n-1)!!}{n!}\left[1+\frac{2n_T+1}{(n_T+1)^2}(Sinhr)^2\right]^{-\frac{2n+1}{2}}\frac{{n_T}^n}{(n_T+1)^{n+1}}$$
$$\times F\left(-\frac{n}{2},-\frac{n-1}{2},\frac{1-2n}{2},\nu^2\right)$$
$$\nu = \left[1-\left(\frac{Sinh(2r)}{Sinh(2r_s)}\right)^2\right]^{1/2}, Im(\nu) \geq 0 \qquad 25$$
$$r_s = \frac{1}{2}ln(2n_T+1)$$

Further, $P_{Sth}(n)$ can also be written as

$$P_{Sth}(n) = \left[1+\frac{2n_T+1}{(n_T+1)^2}(Sinhr)^2\right]^{-\frac{2n+1}{2}}\frac{{n_T}^n}{(n_T+1)^{n+1}}\nu^n P_n\left(\frac{1}{\nu}\right), \qquad 26$$

with $P_n$ being the Legendre polynomial. Eq. 25 was obtained by application of the summation formula on Eq. 10 and 18 for products of two Hypergeometric functions, namely[16]

$$\sum_{n=0}^{\infty} C_\lambda^n s^n F(-n,b,-\lambda,\mu_1)\,F(-n,B,-\lambda,\mu_2) = (1+s)^{\lambda+b+B}(1+s-s\mu_1)^{-b}(1+s-s\mu_2)^{-B}$$
$$\times F\left(b,B,-\lambda,-\frac{s\mu_1\mu_2}{(1+s-s\mu_1)(1+s-s\mu_2)}\right) \qquad 27$$
$$C_\lambda^n = \frac{(-1)^n\Gamma(n-\lambda)}{\Gamma(-\lambda)}$$

A similar expression was given in Ref. (10), Eq. 5.20. After a close comparison of the two Marian expressions, it was found that the results in Ref. (10) contained a small error it that the factor $Cosh(2r_s) + Cosh(2r)$ should have been squared in Eq. 5.20. In Eq. 25, the critical squeezing parameter $r_s$ is defined. The density matrix behaves differently for r greater or less than that value, in that, below the critical value, the even and odd values of the density matrix follow the same curve. Significantly above that value, the even and odd values do not have the same behavior until larger values of n. Near the critical value, the density matrix becomes difficult to calculate and at r=$r_s$ the density matrix is determined directly from Eq. 25 as

$$P_{Sth}(n) = \frac{(2n-1)!!}{n!}\left[1+\frac{2n_T+1}{(n_T+1)^2}(Sinhr)^2\right]^{-\frac{2n+1}{2}}\frac{{n_T}^n}{(n_T+1)^{n+1}};\ \nu = 0\ or\ r = r_s \qquad 28$$

---

[15] Paulina Marian, Higher-order squeezing and photon statistics for squeezed thermal states, Phy. Rev. A, Vol 45, # 3, p. 2044

[16] A. Erdelyi, W. Magnus, F. Oberhettinger, and F. G. Tricomi, Higher Transcendental Functions, (McGraw-Hill, New York, 1953), Vol. 1, Sec. 2.5.2, Eq. (12).

Examples of the density matrix for r below and above the critical value are shown below. Only values on n up to 100 are plotted to show the desired behavior.

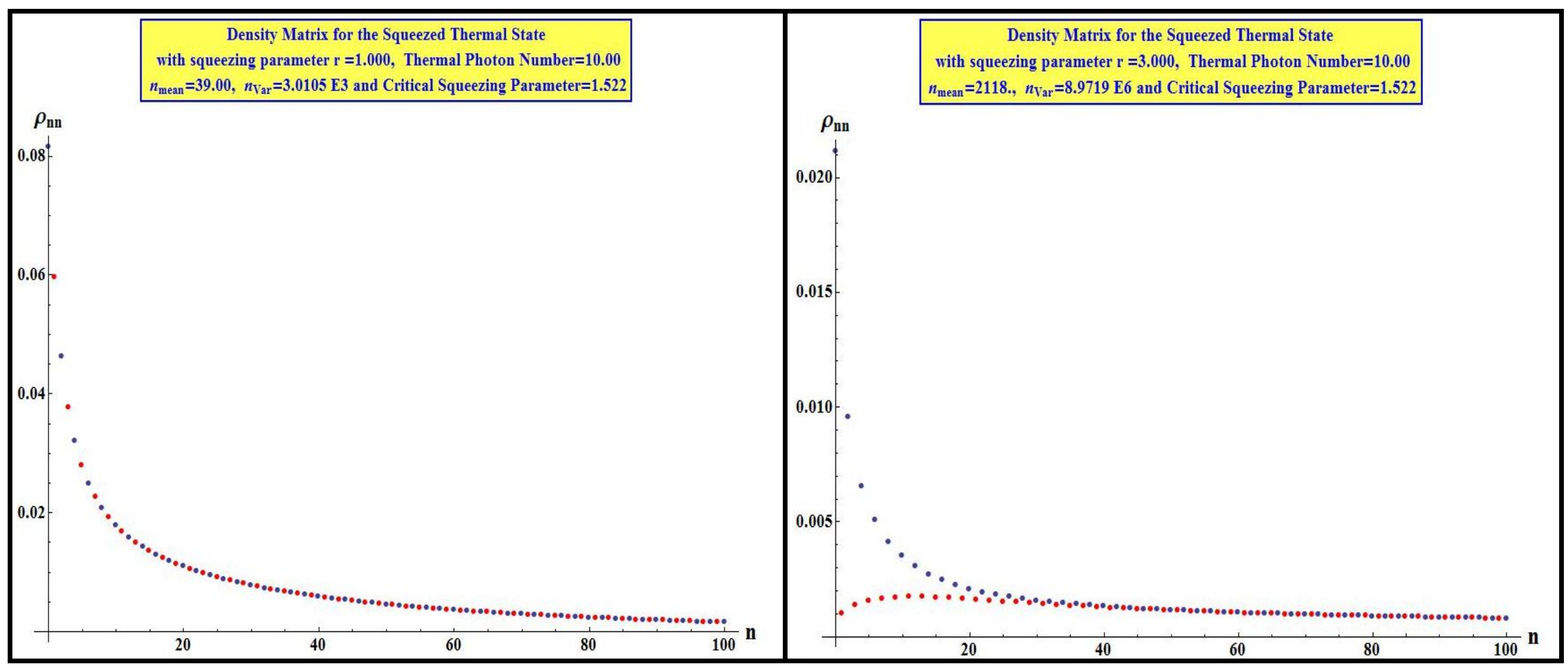

**Table 7: Examples of the density matrix for the squeezing parameter r lesser and greater than the critical squeezing parameter $r_s$**

Examples for the time behavior of the first order correlation function for the squeezed thermal state are presented in the following table. Variables of interest are the mean number of thermal photons $n_T$, the squeezing parameter r and the non-resonance factor $\Delta$.

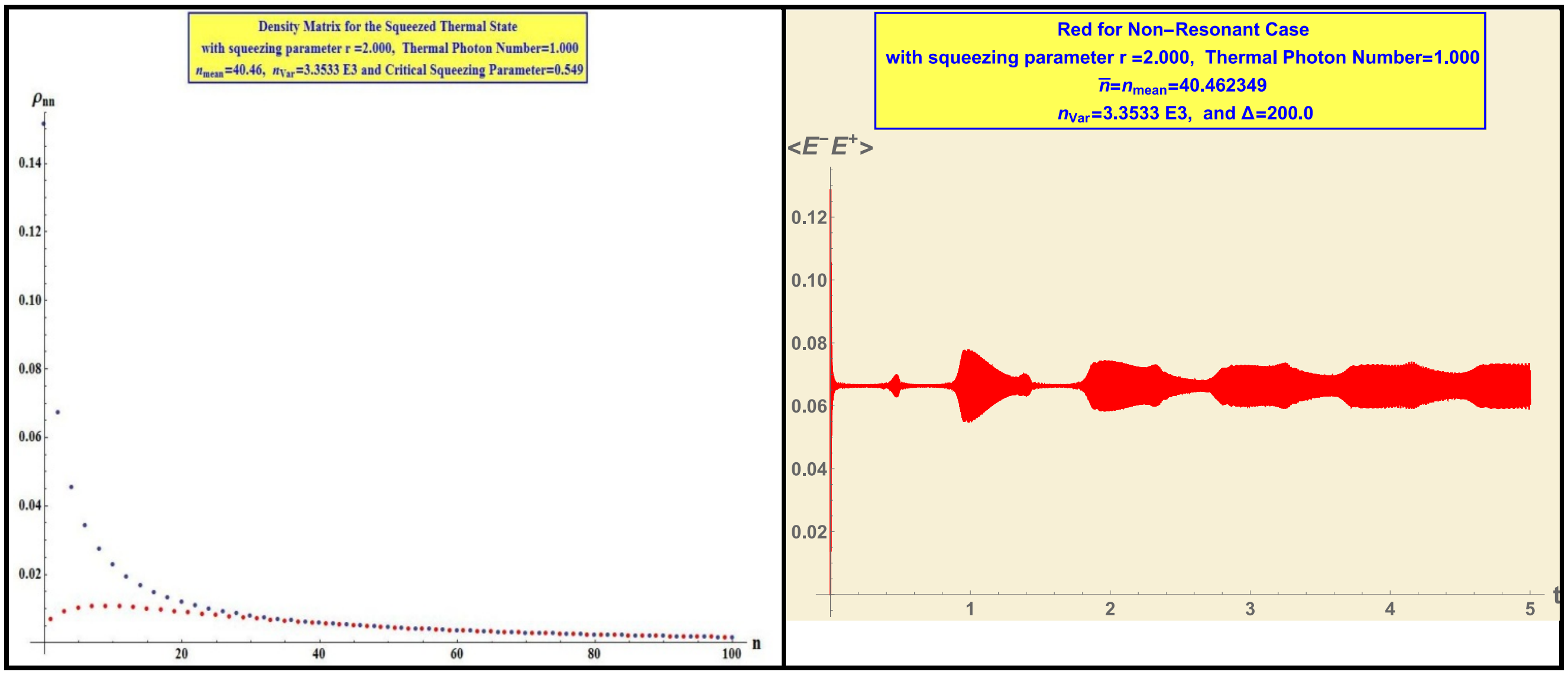

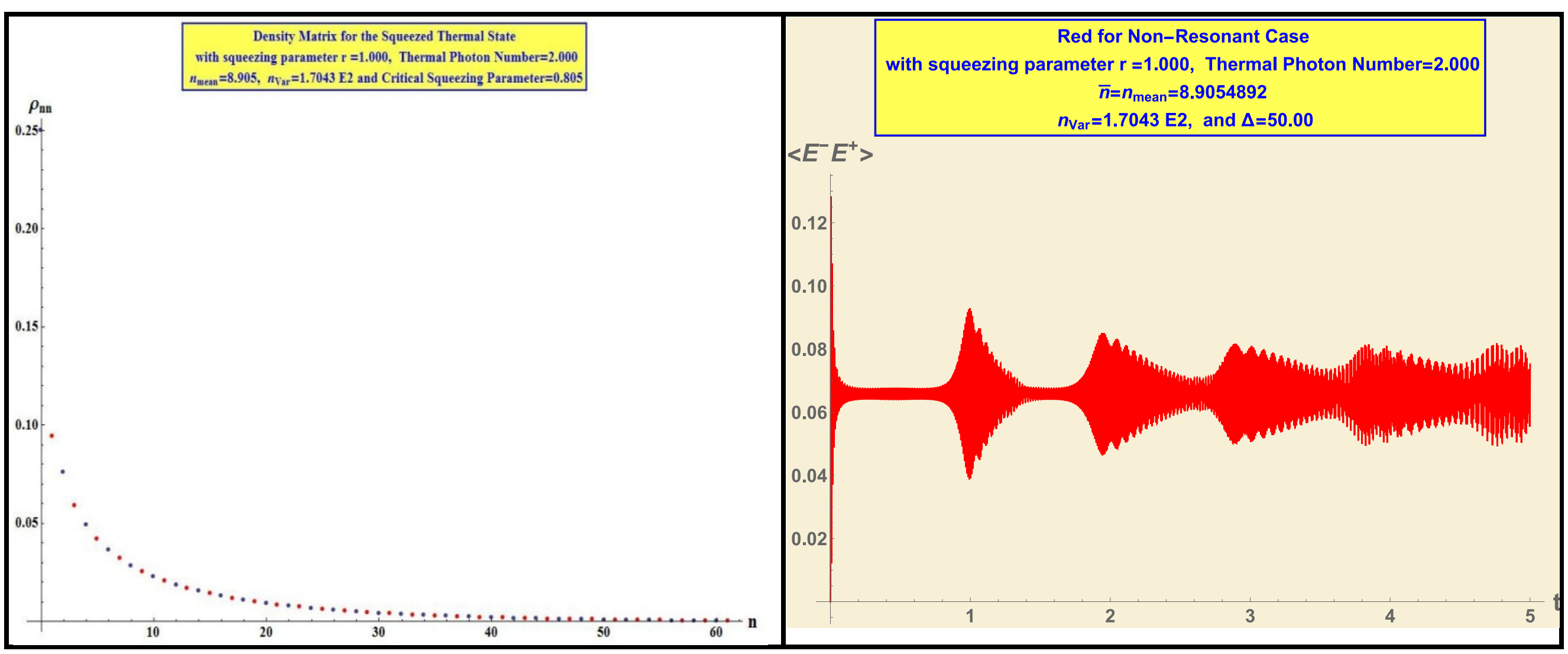

**Table 8: The left-hand columns contain the diagonal elements of the density matrix for the squeezed Thermal. The right-hand columns contain the corresponding $S_1(\overline{n}, \gamma t)_{nr}$ .**

Calculations involving larger values of r and Δ require much larger values of normalized time to display the time behavior.

## Squeezed Coherent State

The density matrix for the squeezed coherent state depends on 3 parameters, namely the total number of coherent photons $|\beta|^2$ where $\beta$ is also called the displacement parameter, the amount of squeezing r, and the total phase ψ. From an examination of Eq. 11, it is seen that the minimum variance occurs for ψ=$\pi$. In addition, there is a minimum in the variance based on r as well such that this variance can be less than for the un-squeezed state. Other interesting cases exist when the phase ψ=$\pi/2$ or 0. The time behavior for the non resonant first order correlation function for the field depends on the amount of non-resonance$\Delta$. As examples see the following table

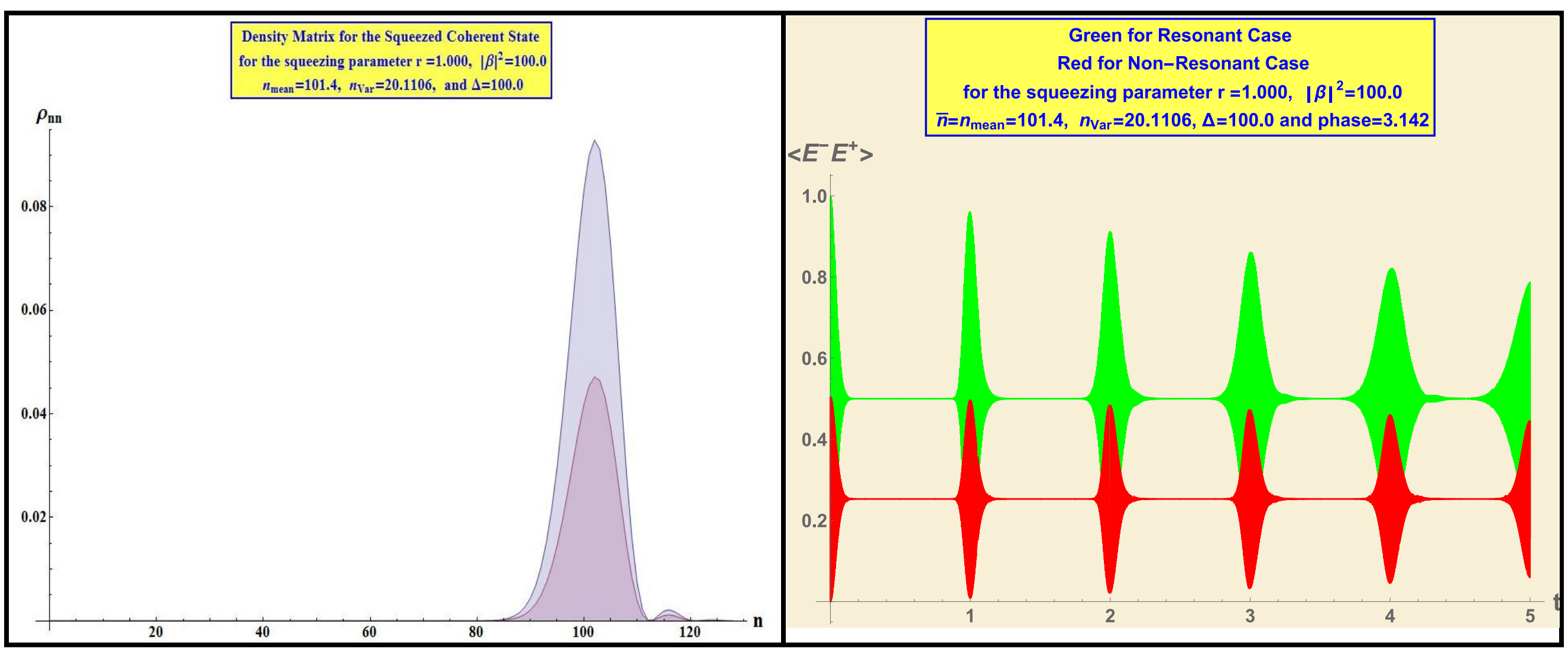

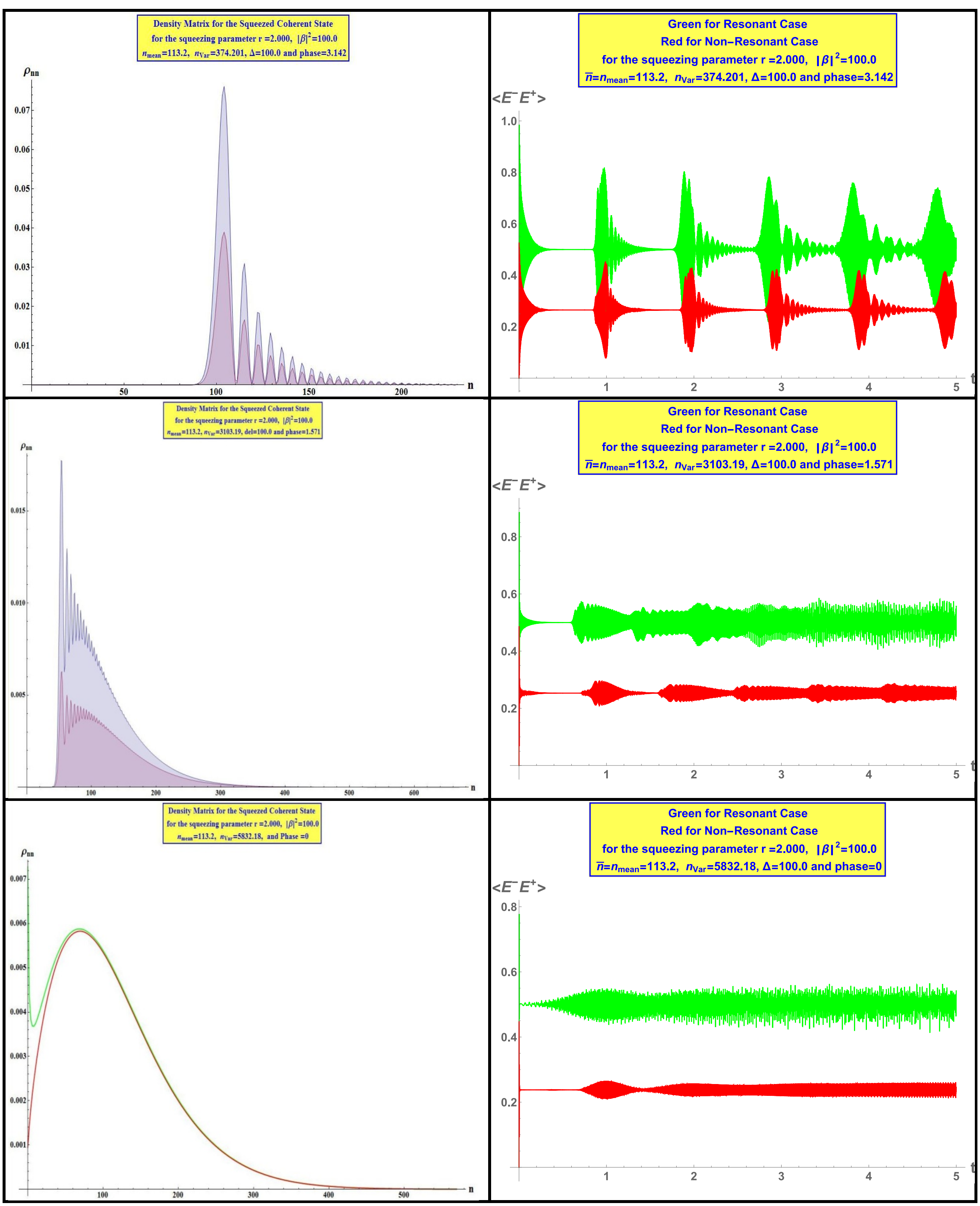

**Table 9: The left-hand columns contain the diagonal elements of the density matrix for the Squeezed Coherent State and for the lower curve the density matrix multiplied by the non-resonance factor in Eq. 2 for the first three examples. In the last example, the density matrix is plotted for the even and odd values of n. The right-hand columns contain the corresponding $S_1(\bar{n},\gamma t)_r$ as the upper curves and $S_1(\bar{n},\gamma t)_{nr}$ for the lower curves.**

More examples were plotted than normal to demonstrate the squeezing provided by r when the phase $\psi = \pi$ for the first two examples. In the upper example, r is small enough to provide a variance of 20 which is a factor of 5 smaller than exits for the pure coherent state. For r greater than the minimum, oscillations in the density matrix quickly set in. This is seen in the second set of curves and note that the first order correlation function for the field replicates these oscillations for both the resonant and non-resonant case. The lower two examples also use a squeezing parameter of 2 but for ψ=$\pi/2$ and 0. For both cases, the density matrix exhibits rapid oscillation for smaller values on n. In fact, the density matrix follows separate curves for the even and odd values of n which is more clearly demonstrated in the 4$^{th}$ set of curves. The third set of curves for the correlation function still exhibits some oscillation corresponding to the density matrix while for last nearly all variation is washed out even for the non-resonant case with a non-resonant factor of 100 with the time variation trending towards the noise state. Of course, for larger values of non-resonance the oscillations would continue out to larger values of normalized time.

## Mixed Squeezed Coherent State with the Thermal State

Vourdas[9] used the Glauber P representation[6] to define the density matrix or the mixed squeezed coherent state with thermal noise in the squeezed coherent basis. He then found the coherent state diagonal matrix elements of this density matrix and used this as a generating function to determine the number state diagonal elements of the density matrix. As stated in the footnotes, he did not define the mean or variance; however, he stated that there was a one to one correspondence between the coherent states and the squeezed coherent states. Using this correspondence, the mean number of photons is then the sum of the squeezed coherent state mean photon number and thermal state number of photons. In addition, the variance is found as in the case of the mixed coherent and thermal states as the sum of the squeezed coherent state variance, the thermal state variance and the product of twice the means of the squeezed coherent state and thermal state. In Appendix A, an alternative expression for the matrix elements of the density matrix is provided.

Only two examples are shown. The first is for an equal number of coherent and thermal photons and a squeezing parameter on 1. The second case is for a large number of coherent photons and a small mean average of thermal photons for a squeezing parameter of 2. Recall that the derivation given was for the case of the smallest variance in photon number. Even for the second case, there is no oscillation seen in the density matrix as discussed above. For the first case, the first order correlation function quickly degrades since the non-resonance factor is only 50.

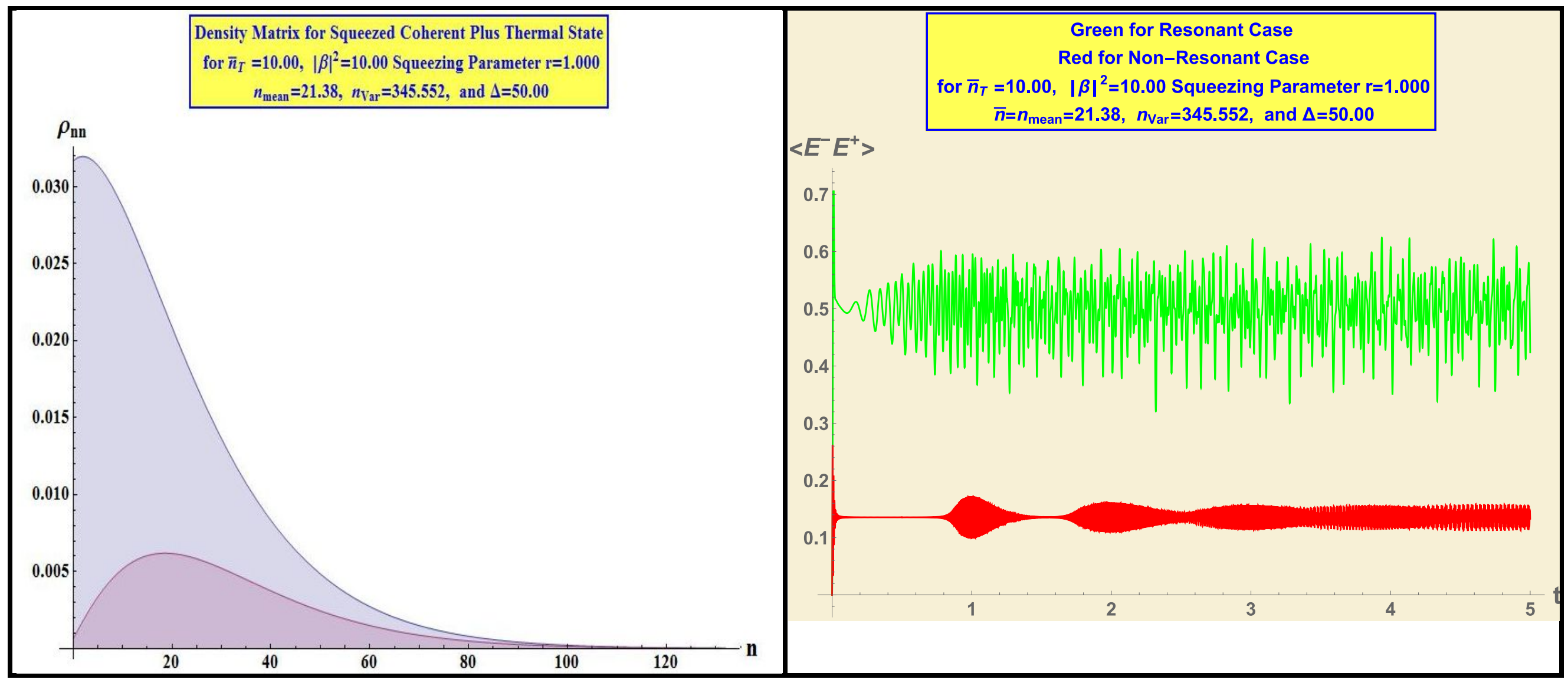

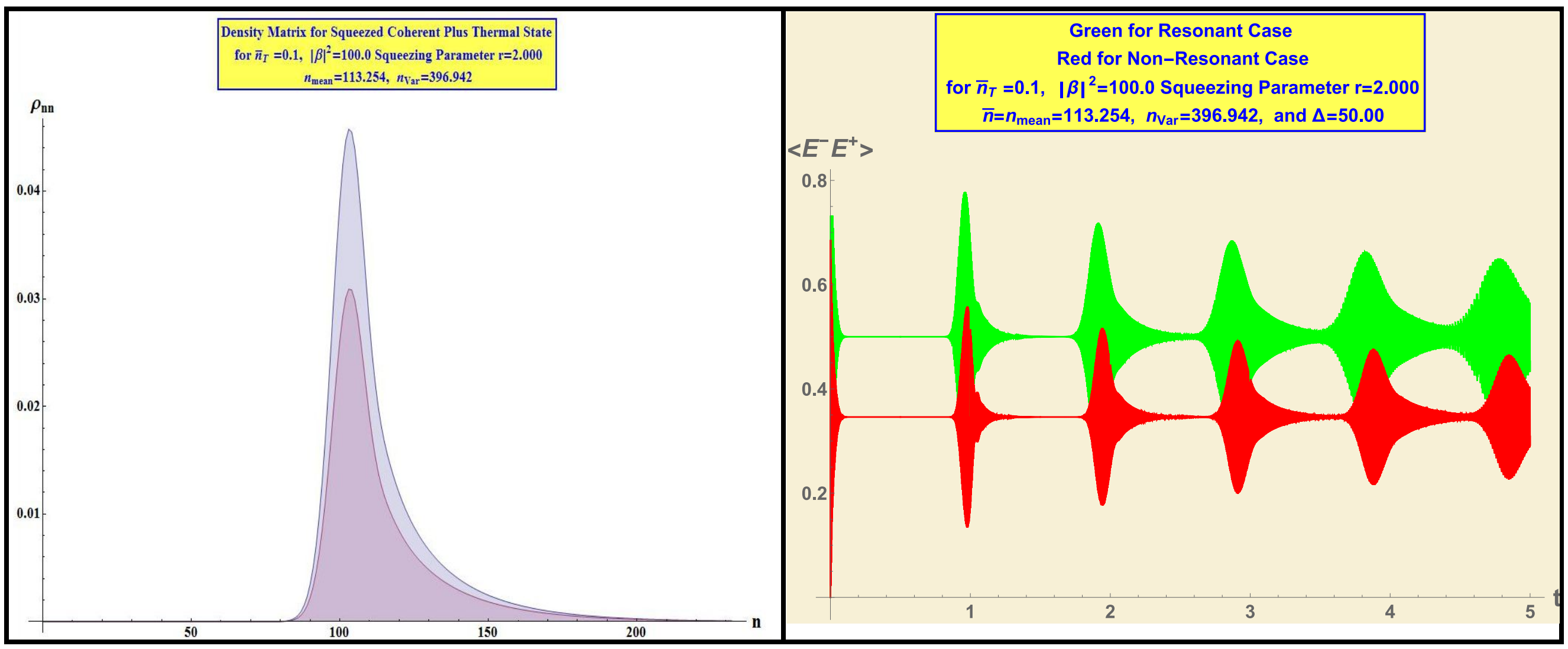

**Table 10: The left-hand columns contain the diagonal elements of the density matrix for the squeezed coherent state mixed with a thermal state and for the lower curve the density matrix multiplied by the non-resonance factor in Eq. 2. The right-hand columns contain the corresponding $S_1(\bar{n},\gamma t)_r$ as the upper curves and $S_1(\bar{n},\gamma t)_{nr}$ for the lower curves.**

If one compares the density matrix in the second example for the squeezed coherent state, one can easily see the effect of mixing even for a small number of thermal photons is that there are no oscillations in the density matrix or sub-oscillations in the correlation functions.

## Displaced Squeezed Thermal State

The displaced squeezed thermal state (DSTS) is unlike the previous case in that displacement and squeezing applies the thermal state unlike the previous case which can be thought of as a squeezed displaced vacuum state mixed with the thermal state. That is, there was no displacement or squeezing on the thermal state as there is for the current example.

Three examples are provided. In the first example, there is a relatively large mean thermal photon number and it is seen that the density matrix contains no oscillation for a squeezing parameter of 1. Oscillations will occur for squeezing larger than 2. If the mean number of thermal photons is reduced, oscillations begin at much smaller values of the squeezing parameter. The second two examples should be compared to those provided in table 9 for the squeezed coherent state. It is seen that even for the small number of thermal photons, the density matrix broadens and the first order correlation functions begins to decay sooner. The effect is more pronounced for the squeezing parameter of 2.

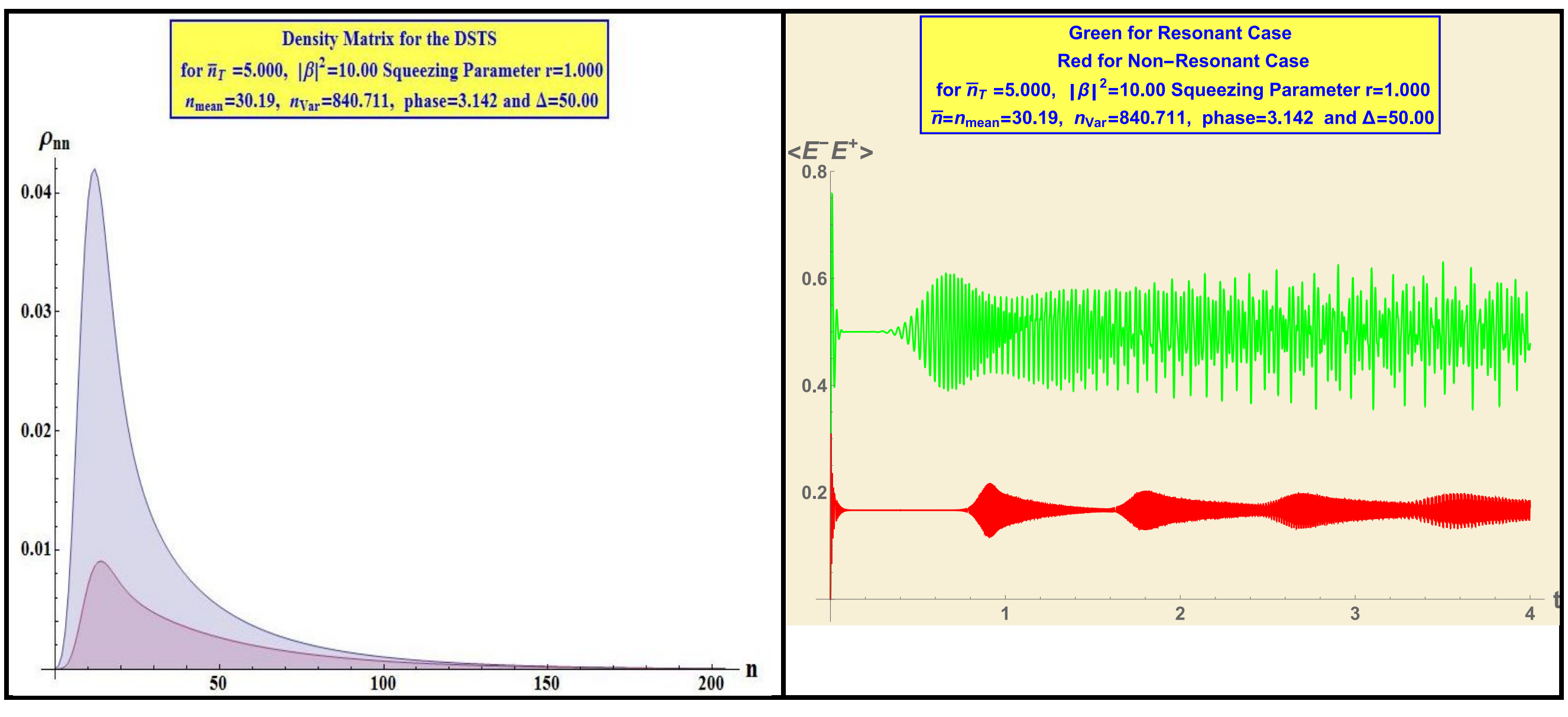

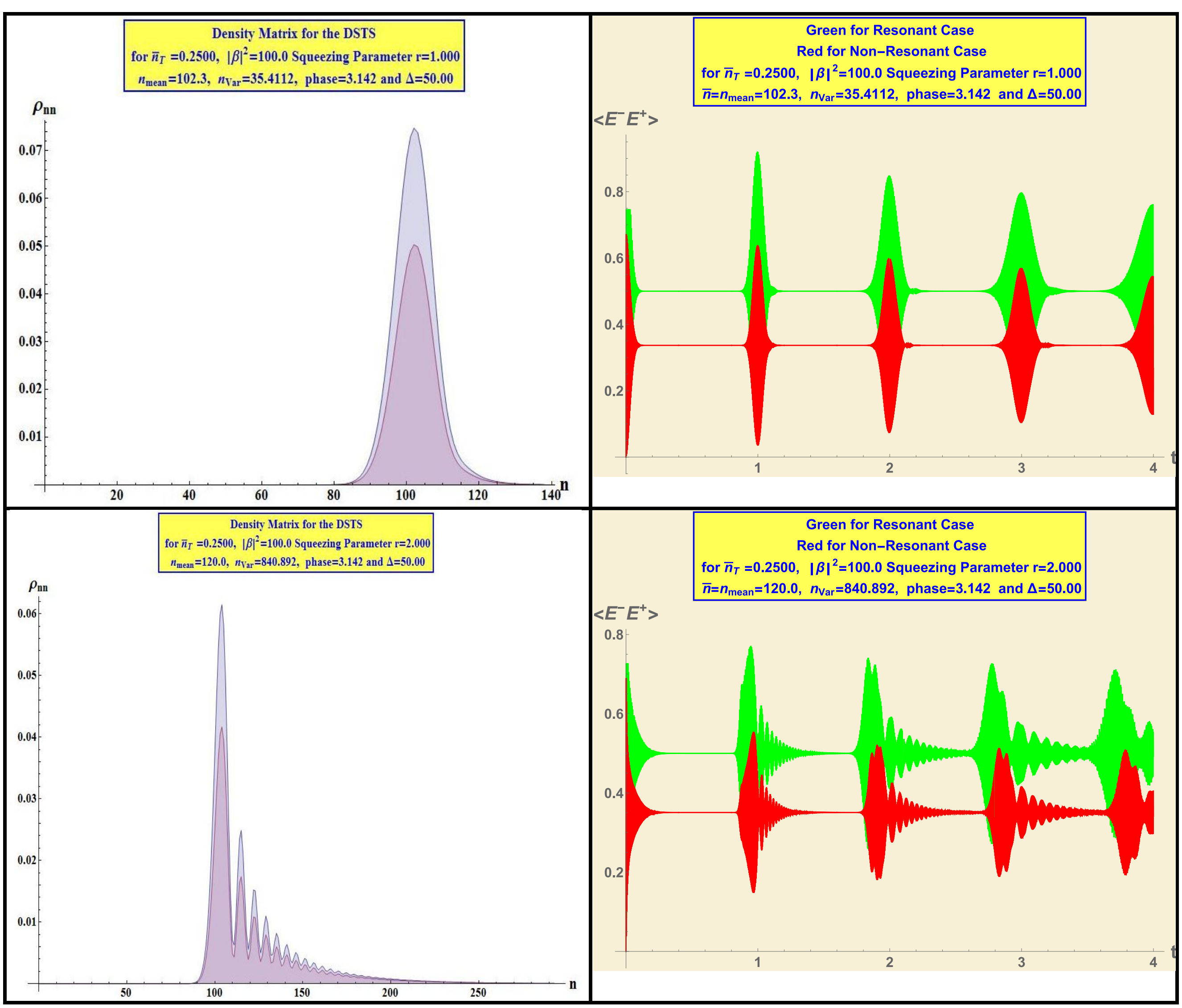

**Table 11: The left-hand columns contain the diagonal elements of the density matrix for the DSTS and for the lower curve the density matrix multiplied by the non-resonance factor in Eq. 2. The right-hand columns contain the corresponding $S_1(\bar{n},\gamma t)_r$ as the upper curves and $S_1(\bar{n},\gamma t)_{nr}$ for the lower curves.**

## Displaced Number State

The displaced number state is easily found from the straightforward application of the displacement operator on the number state. The density matrix resembles the higher-level states of the harmonic oscillator.

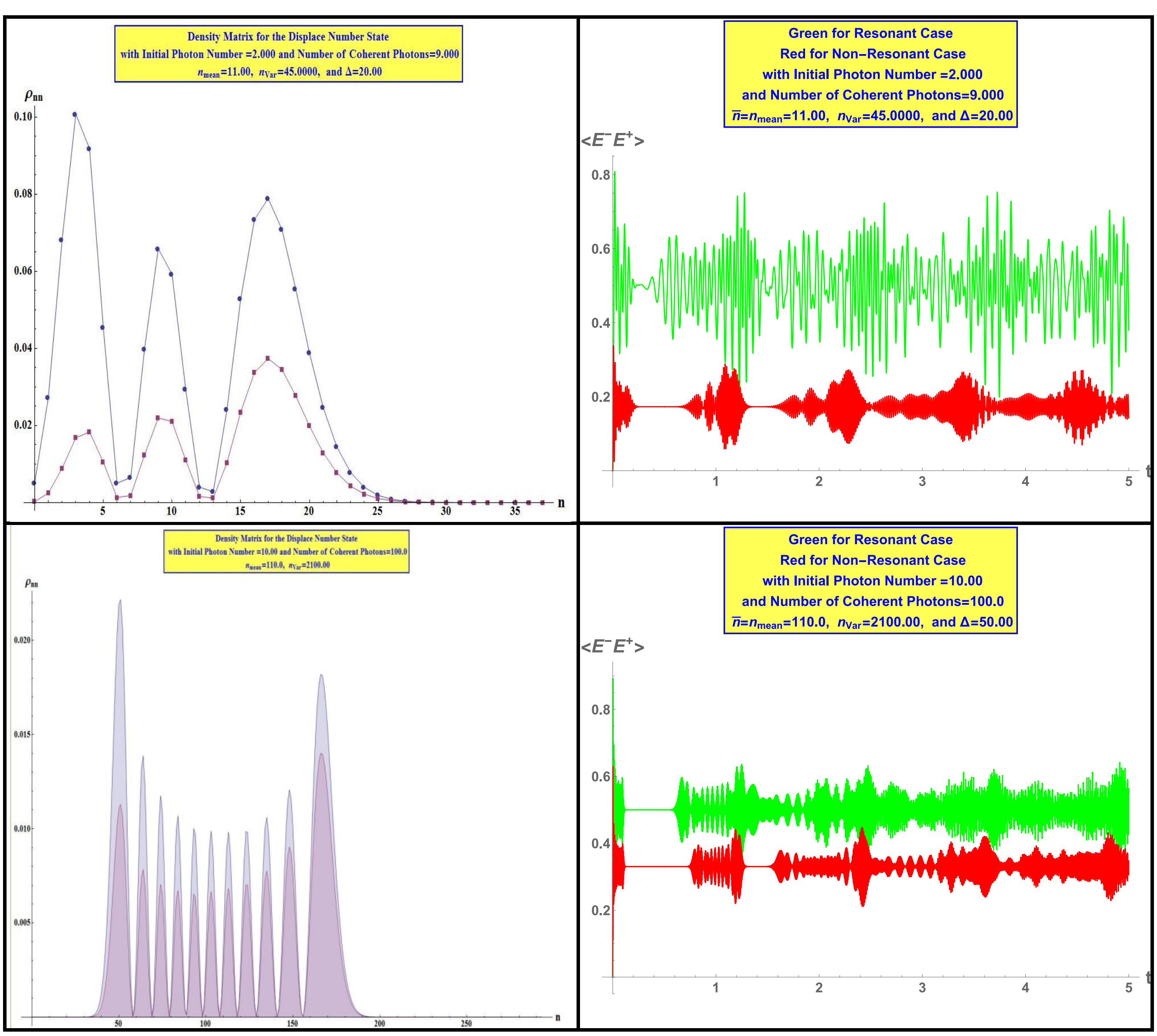

**Table 12: The left-hand columns contain the diagonal elements of the density matrix for the displaced number states and for the lower curve the density matrix multiplied by the non-resonance factor in Eq. 2. The right-hand columns contain the corresponding $S_1(\bar{n},\gamma t)_r$ as the upper curves and $S_1(\bar{n},\gamma t)_{nr}$ for the lower curves.**

If the amount of displacement (or the mean number of coherent photons) is less than ½ the initial photon number, the density matrix is compressed on the left and the number of oscillations falls below the photon number. The correlation function again contains oscillations resembling the oscillations in the density matrix with the non-resonant case decaying more quickly.

## Squeezed Displaced Number State

The squeezed displaced number state (SDNS) was calculated using Eq. 15. The calculation of the density matrix was not immediately straightforward in Mathematica since there was an issue of taking the sum of the squares of the real and imaginary parts of the summation which in most cases had hugely different degrees of accuracy which resulted in an incorrect result of 0 for larger values on n. The problem was eventually solved by multiplying by the normalization coefficient before performing the sum, although a longer calculation time resulted. This difficulty also led to an exploration of alternate formulations. Two alternate expressions are provided in Appendix B.

The number of possible examples is large since the displacement (or number of coherent photons), the number state photon number, the squeezing parameter, the off resonance parameter and the overall phase can be varied. Only a few examples are provided.

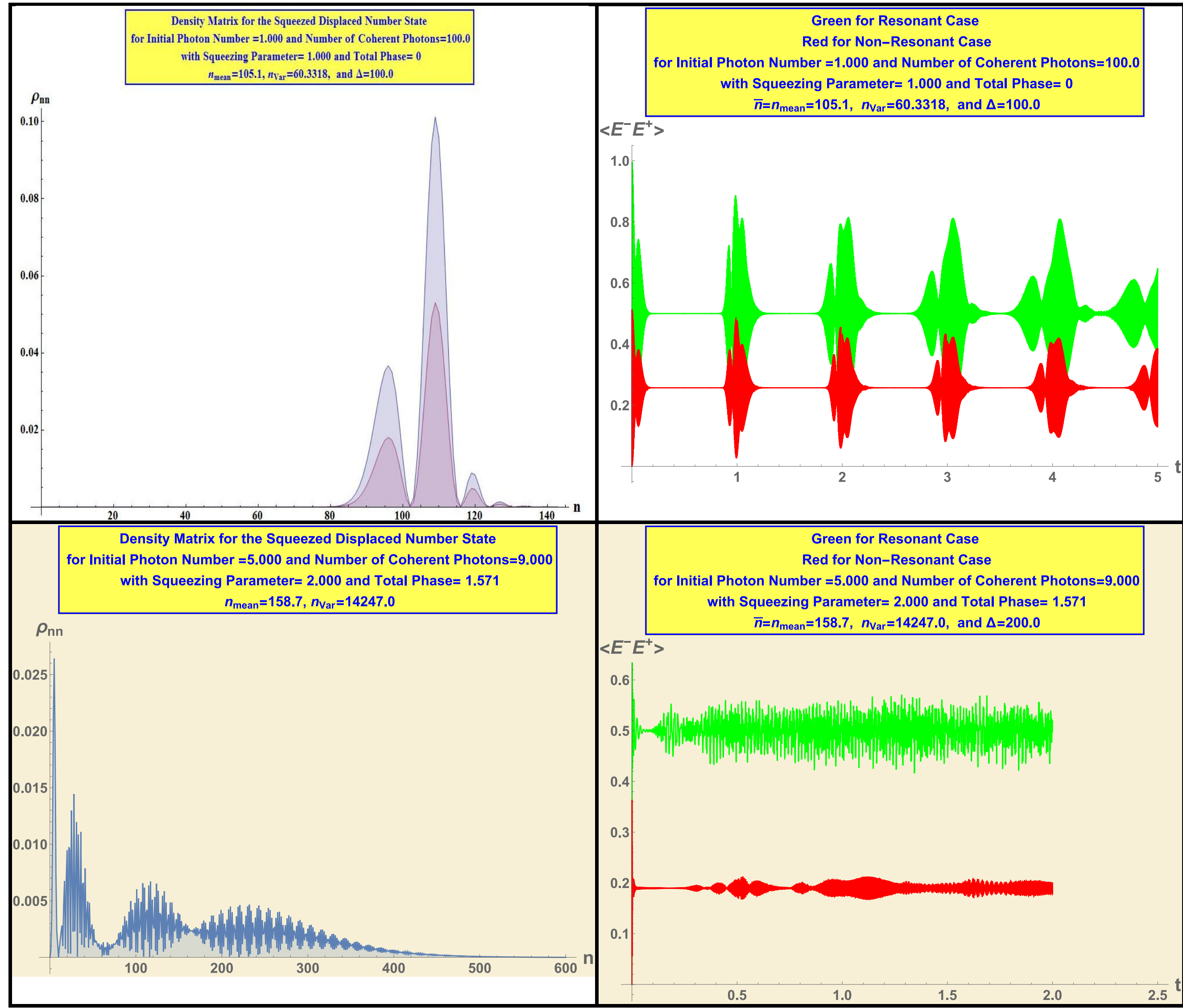

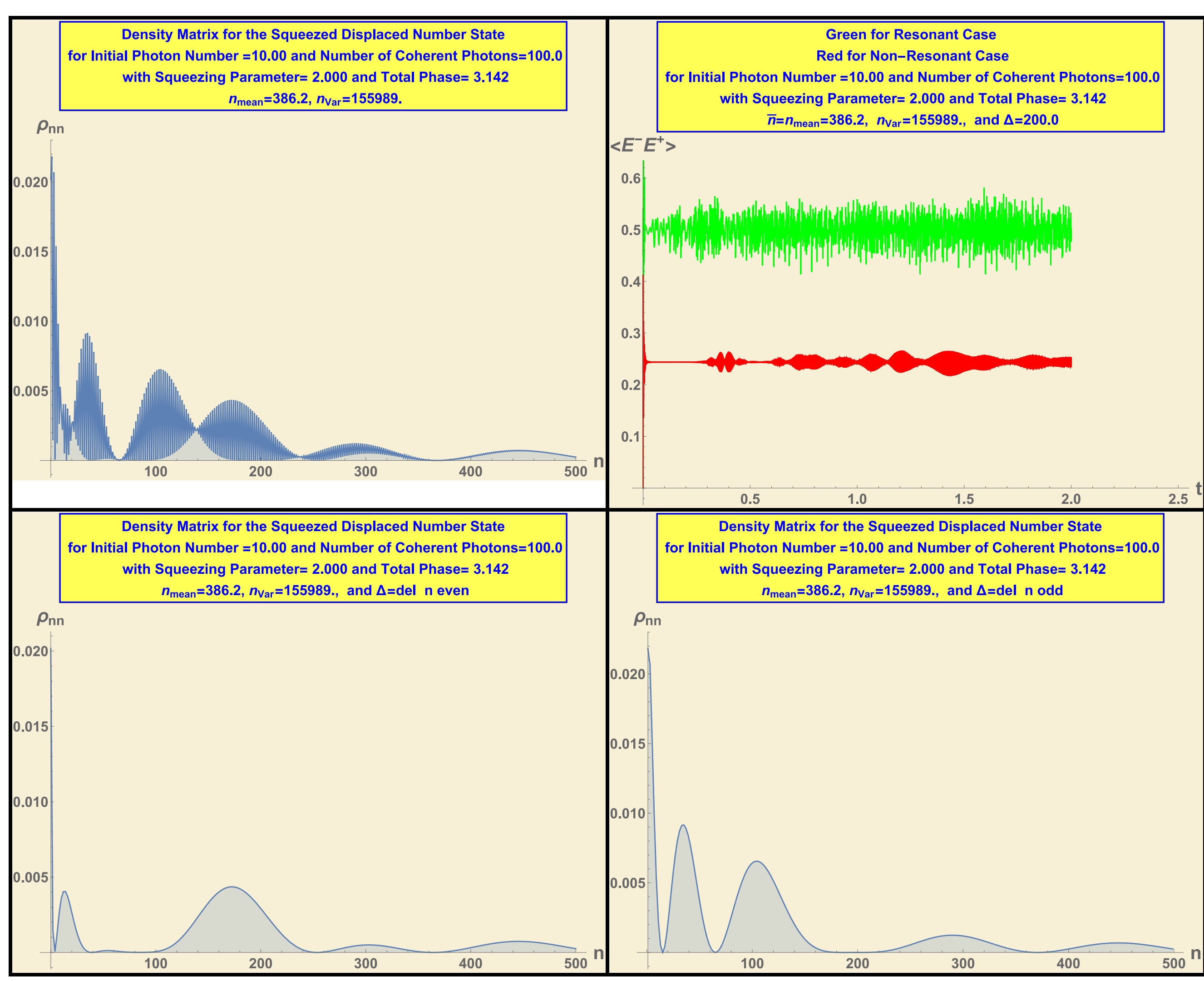

**Table 13: The left-hand columns contain the diagonal elements of the density matrix for the squeezed displaced number states and for the first example the lower curve the density matrix multiplied by the non-resonance factor in Eq. 2. The right-hand columns contain the corresponding $S_1(\bar{n}, \gamma t)_r$ as the upper curves and $S_1(\bar{n}, \gamma t)_{nr}$ for the lower curves.**

As in previous cases, non-resonance provides Rabi type oscillations to greater values of normalized time. For the first example, a squeezing parameter of 1 and phase of 0 was used to minimize the variance. For the second two examples a squeezing parameter of 2 was use while the phase of $\frac{\pi}{2}$ and $\pi$ were used since they provided interesting examples. For the last example, the even or odd parts of the density matrix were plotted which results in very smooth curves. This shows that the rapid variation shown when every value of n is plotted is due to the difference in even and odd values of n. This is apparent in the dark and light portions of the density matrix oscillating rapidly between even and odd values of the density matrix. As an aside, it is probably possible to express the DSTS as an infinite sum of SDNS as was done in a similar fashion for the squeezed thermal state being a sum of squeezed number states.

## Observations

Photon echo revivals do indeed fall at the revival time predicted by Erberly, namely $\tau_r = 2\,\pi\,\frac{\sqrt{\bar{n}+1+\Delta}}{\gamma}$ for all photon densities except those that only have values for every other photon number (See for every case considered except Table 4 and Table 6 above.). For those cases revivals fall at one half the revival time.

The Squeezed Displaced Number State exhibits some unusual behavior. For a squeezing parameter of one, shown in the top row of Table 13, both the resonant and non-resonant values of the peaks of the revivals ($S_1(\bar{n},\gamma t)_r$ *and* $S_1(\bar{n},\gamma t)_{nr}$) fall at the integer revival time. On the other hand, the second two rows (for a squeezing parameter of 2) fall at values which to not exactly match ether ½ or integer values of revival time. The photon densities have values which are not zero. Instead the photon density fluctuates greatly. The resonant values for the squeezing parameter of 2 quickly degenerated to noise.

The photon revivals often show a considerable spread particularly when the photon density distribution is considerable different for even or odd values of n. It should be noted that the revivals for resonance and non-resonance do not in fact fall at the same value of time but appear to since both are plotted with respect to the corresponding revival time. The inclusion of non-resonance has resulted in Rabi type oscillations which exist for a considerably longer time than for the resonant case. This occurs even for the pure thermal state and for cases of significant thermal components for mixed states. It appears that the suppression of the smaller photon numbers in the calculation for $S_1(\bar{n},\gamma t)_{nr}$ leads to these results. Thus, it appears that only first order coherence is required to produce oscillation in the case of non-resonance.

It should be noted that in several places within this document the phrase "to much longer times" was used. By using the Eberly[3] photon revival time, that phrase loses its meaning. The phrases now means to times longer that 10-15 revival periods.

## Appendix A

There is another expression for the matrix elements of the Mixed Squeezed Coherent State with the Thermal State that may be of interest. One could express this directly using the definitions in Ref. 10 as

$$\langle n|\hat{\rho}|m\rangle = \frac{1}{\pi\cosh(r)\langle n_T\rangle\sqrt{n!\,m!}}\int Exp\left\{-\frac{1}{\langle n_T\rangle}\left[e^{-2r}(A_R-A_0)^2+e^{2r}A_I^2\right]-|A|^2+\frac{1}{2}\left(A^2+(A^*)^2\right)tanh(r)\right\}$$
$$\times\left[\frac{1}{2}tanh(r)\right]^{\frac{n+m}{2}} H_n\left\{\frac{A}{[sinh(2r)]^{\frac{1}{2}}}\right\} H_m^*\left\{\frac{A}{[sinh(2r)]^{\frac{1}{2}}}\right\} d^2A, \qquad \text{A1}$$

assuming that $\theta = \lambda = 0$ and

$$\hat{\rho} = \int P(A)|A, r\theta\lambda >< A, r\theta\lambda|\, d^2A$$

$$\langle n|A; r\theta\lambda\rangle = \frac{exp\left(-\frac{1}{2}|A|^2+\frac{1}{2}e^{-i\vartheta}A^2 tanh(r)\right)\left[\frac{1}{2}e^{i\vartheta}tanh(r)\right]^{\frac{n}{2}}}{[\cosh(r)]^{\frac{1}{2}}\sqrt{n!}} H_n\left\{\frac{A}{[e^{i\vartheta}sinh(2r)]^{\frac{1}{2}}}\right\} \qquad \text{A2}$$

The integration in Eq. 29 can be performed as letting A=$A_R + iA_I$ and

$$H_n(x+y) = \sum_{k=0}^{n}\binom{n}{k} H_k(x)(2y)^{n-k} = 2^{-\frac{n}{2}}\sum_{k=0}^{n} H_{n-k}\left(\sqrt{2}x\right) H_k\left(\sqrt{2}y\right) \qquad \text{A3}$$

Integrating over $A_R$ and $A_I$ separately yields

$$\begin{aligned}
\langle n|\hat{\rho}|m\rangle = {} & \frac{Norm C_1 C_2}{\sqrt{n!\,m!}}\left[\frac{1}{2}tanh(r)\right]^{\frac{n+m}{2}} Exp\left[-\frac{e^{-\rho}A_0^2\langle n_T\rangle}{\langle n_T\rangle\left[\langle n_T\rangle + e^{-2r}cosh^2(r)\left(1+tanh(r)\right)\right]}\right] \\
& \times \sum_{k=0}^{n}\sum_{k'=0}^{m}\left\{\frac{2i}{C_1\sqrt{sinh(2r)}}\right\}^{n-k}\left[\frac{-2i}{C_1\sqrt{sinh(2r)}}\right]^{m-k'}\begin{pmatrix} 0 & n+m-k-k' \ odd \\ \Gamma\left(\frac{n+m-k-k'+1}{2}\right) & n+m-k-k' \ even \end{pmatrix} \\
& \times \sum_{l=0}^{k}\sum_{l'=0}^{k'} H_l\left(\frac{e^{-2r}A_0}{\langle n_T\rangle\sqrt{sinh(2r)}}\right) H_{l'}\left(\frac{e^{-2r}A_0}{\langle n_T\rangle\sqrt{sinh(2r)}}\right)\binom{k}{l}\binom{k'}{l'}\left(\frac{2}{C_2\sqrt{sinh(2r)}}\right)^{k-l+k'-l'} \\
& \times \begin{pmatrix} 0 & k+k'-l-l' \ odd \\ \Gamma\left(\frac{k+k'-l-l'+1}{2}\right) & k+k'-l-l' \ even \end{pmatrix}
\end{aligned} \qquad \text{A4}$$

$$\begin{aligned}
Norm &= \frac{1}{\pi\cosh(r)\langle n_T\rangle} \\
C_1 &= \sqrt{\frac{\langle n_T\rangle}{e^{2r} + \langle n_T\rangle\left(1+tanh(r)\right)}} \\
C_2 &= \sqrt{\frac{\langle n_T\rangle}{e^{-2r} + \langle n_T\rangle\left(1-tanh(r)\right)}}
\end{aligned} \qquad \text{A5}$$

If one were to compare this with the Vourdas[(9)] expression (26), an expansion for the generalized Hermite Polynomial is seen. This however is not used to calculate the examples above.

## Appendix B

Dantus[17] and Kim[18] have also obtained expressions for the density matrix of the Squeezed Displaced Number State. In particular it can be found that the squeezed displaced state can be found as a infinite sum of the product of the matrix elements in the number representation of the squeezing operator and of the displacement operator. Unfortunately, Dantus's expression 2.11 does not account for terms for m<n. In Appendix B, two alternate expressions for the density matrix are presented

[17] Célia M. A. Dantas, Norton G. de Almeida and B. Baseia, Statistical Properties of the Squeezed Displaced Number States, Brazilian Journal of Physics, vol. 28, # 4, p.462.

[18] Sang Pyo Kim, Time-Dependent Displaced and Squeezed Number States, Journal of the Korean Physical Society, Vol. 44, # 2, p. 446

Instead we write

$$\rho_{nn}(m) = N_1 \frac{|\gamma|^2}{Cosh^2(r)} \left[\frac{Tanh(r)}{2}\right]^{-1} \begin{matrix} Sum(even) & n\ even \\ Sum(odd) & n\ odd \end{matrix} \qquad \text{B1}$$

In the previous expression

$$Sum(even) = \left| \sum_{j=0}^{\infty} \gamma^{2j} (2j)!\,(-1)^j Exp[-j\theta I] \left[\frac{Tanh(r)}{2}\right]^j \right.$$
$$\left. \times \left\{ \sum_{k=0}^{Min[m,2j]} \frac{(-1)^k |\gamma|^{-2k}}{k!\,(m-k)!\,(2j-k)!} \right\} \left\{ \sum_{p=0}^{Min\left[\frac{n}{2},j\right]} \frac{\left[\frac{-4}{Sinh^2(r)}\right]^p}{(2p)!\,(j-p)!\left(\frac{n}{2}-p\right)!} \right\} \right|^2$$

$$Sum(odd) = \left| \sum_{j=0}^{\infty} \gamma^{2j} (2j+1)!\,(-1)^j Exp[-jI\theta] \left[\frac{Tanh(r)}{2}\right]^j \right. \qquad \text{B2}$$
$$\left. \times \left\{ \sum_{k=0}^{Min[m,2j+1]} \frac{(-1)^k |\gamma|^{-2k}}{k!\,(m-k)!\,(2j+1-k)!} \right\} \left\{ \sum_{p=0}^{Min\left[\frac{n-1}{2},j\right]} \frac{\left[\frac{-4}{Sinh^2(r)}\right]^p}{(2p+1)!\,(j-p)!\left(\frac{n-1}{2}-p\right)!} \right\} \right|^2$$

$$N_1 = \frac{m!\,n!}{Cosh(r)} (|\gamma|^2)^m e^{-|\gamma|^2};\ \gamma = Cosh(r)\beta + Sinh(r) e^{I\theta} \beta^*$$

An alternative expression can be obtained directly from the expression for the squeezed displaced number state, namely

$$|\gamma, m>_g = S(\zeta)D(\gamma)|m> = S(\zeta)D(\gamma)\frac{\left(b^\dagger\right)^m}{\sqrt{m!}}|0> = S(\zeta)\frac{\left(b^\dagger - \gamma^*\right)^m}{\sqrt{m!}}D(\gamma)|0>$$
$$= \frac{\left(g^\dagger - \gamma^*\right)^m}{\sqrt{m!}} S(\zeta)D(\gamma)|0> = \sum_{i=0}^{m} \binom{m}{i} \frac{\left(g^\dagger\right)^{m-i} (-\gamma^*)^i}{\sqrt{m!}} S(\zeta)D(\gamma)|0>. \qquad \text{B3}$$

Then

$$\langle n|\gamma, m\rangle_g = \sum_{i=0}^{m} \binom{m}{i} \frac{(-\gamma^*)^i}{\sqrt{m!}} \sum_{k=0}^{\infty} \langle n|\left(g^\dagger\right)^{m-i}|k\rangle \langle k|S(\zeta)D(\gamma)|0\rangle. \qquad \text{B4}$$

The matrix element for the squeezed displaced vacuum is known.

$$\langle p|S(\zeta)D(\gamma)|0\rangle = \frac{exp\left(-\frac{1}{2}|\gamma|^2 + \frac{1}{2}e^{-i\theta}\gamma^2 tanh(r)\right)}{[\cosh(r)]^{\frac{1}{2}}} \frac{\left[\frac{1}{2}e^{i\theta} tanh(r)\right]^{\frac{p}{2}}}{\sqrt{p!}} H_p\left\{\frac{\gamma}{\left[e^{i\theta} sinh(2r)\right]^{\frac{1}{2}}}\right\} \qquad \text{B5}$$

Using the expression for the normally ordered creation and destruction operators[19]

$$
\begin{aligned}
\langle n|(g^{\dagger})^{m-i}|k'\rangle = & \sum_{k=0}^{Floor\left(\frac{m-i}{2}\right)} \sum_{s=0}^{m-i-2k} \left(\frac{C}{2}\right)^{k} \\
& \times \frac{(m-i)!\,[Cosh(r)]^{s}[Sinh(r)]^{m-i-s-2k}\langle n|(\beta^{\dagger})^{s}(\beta)^{m-i-s-2k}|k'\rangle Exp[i(m-i-s-2k)\theta]}{k!\,s!\,(m-i-2k-s)!}
\end{aligned}
\qquad \text{B6}
$$

and

$$
\langle n|(a^{\dagger})^{s}(a)^{m-i-s-2k}|k'\rangle = \frac{\sqrt{n!\,(\mathrm{n+m-i-2s-2k})!}}{(n-s)!}\delta_{n+m-i-2s-2k,k'} \qquad \text{B7}
$$

Putting the terms all together yields

$$
\begin{aligned}
&\rho_{nn}(m) = \left|\langle n|\beta,m\rangle_{g}\right|^{2} \\
&= \frac{exp\{-|\beta|^{2}[1+cos(\theta+2\varphi)tanh(r)]\}m!\,n!\left[\frac{1}{2}tanh(r)\right]^{n}(|\gamma^{2}|)^{m}}{cosh(r)} \\
&\times \left|\sum_{j=0}^{m}\frac{(-\gamma^{*})^{-j}[Sinh(r)]^{j}}{(m-j)!}[2tanh(r)]^{-\frac{\mathrm{j}}{2}}e^{\frac{(j)I\theta}{2}}\sum_{q=0,1}^{j}\frac{[tanh(r)]^{q}}{\left(\frac{j-q}{2}\right)!}\sum_{s=0}^{Min[q,n]}\frac{2^{s}}{[tanh(r)]^{2s}s!\,(q-s)!\,(n-s)!}\right. \\
&\times \left. H_{\mathrm{n+q-2s}}\left\{\frac{\gamma}{[e^{-i\theta}sinh(2r)]^{\frac{1}{2}}}\right\}\right|^{2}
\end{aligned}
\qquad \text{B8}
$$

---

[19] R. M. Wilcox, J. Math. Phys. 8, 962 (1967)